\documentclass[12pt,a4paper]{amsart}
\usepackage{amsmath,amstext,amssymb,amsthm,amsfonts}
\newcommand{\nc}{\newcommand}
\nc{\one}{\mbox{\bf 1}}
\nc{\invtensor}{\underset{\leftarrow}{\otimes}}
\nc{\const}{\operatorname{const}}

\nc{\ad}{\operatorname{ad}}
\nc{\tr}{\operatorname{tr}}
\nc{\tp}{\operatorname{top}}
\nc{\rank}{\operatorname{rank}}
\nc{\corank}{\operatorname{corank}}
\nc{\codim}{\operatorname{codim}}
\nc{\sdim}{\operatorname{sdim}}
\nc{\mult}{\operatorname{mult}}

\nc{\spn}{\operatorname{span}}
\nc{\Sym}{\operatorname{Sym}}
\nc{\sym}{\operatorname{sym}}
\nc{\id}{\operatorname{id}}
\nc{\Id}{\operatorname{Id}}
\nc{\Ree}{\operatorname{Re}}

\nc{\htt}{\operatorname{ht}}
\nc{\Ker}{\operatorname{Ker}}
\nc{\rker}{\operatorname{rKer}}
\nc{\im}{\operatorname{Im}}
\nc{\osp}{\mathfrak{osp}}
\nc{\sgn}{\operatorname{sgn}}
\nc{\F}{\operatorname{F}}
\nc{\Mod}{\operatorname{Mod}}
\nc{\Mat}{\operatorname{Mat}}
\nc{\Soc}{\operatorname{Soc}}
\nc{\Inj}{\operatorname{Inj}}
\nc{\Hom}{\operatorname{Hom}}
\nc{\End}{\operatorname{End}}
\nc{\supp}{\operatorname{supp}}
\nc{\Card}{\operatorname{Card}}
\nc{\Ann}{\operatorname{Ann}}
\nc{\Ind}{\operatorname{Ind}}
\nc{\Coind}{\operatorname{Coind}}
\nc{\wt}{\operatorname{wt}}
\nc{\ch}{\operatorname{ch}}
\nc{\Stab}{\operatorname{Stab}}
\nc{\Sch}{{\mathcal S}\mbox{\em ch}}
\nc{\Irr}{\operatorname{Irr}}
\nc{\Spec}{\operatorname{Spec}}
\nc{\Prim}{\operatorname{Prim}}
\nc{\Aut}{\operatorname{Aut}}
\nc{\Ext}{\operatorname{Ext}}

\nc{\Fract}{\operatorname{Fract}}
\nc{\gr}{\operatorname{gr}}
\nc{\deff}{\operatorname{def}}
\nc{\HC}{\operatorname{HC}}

\nc{\red}{\operatorname{red}}
\nc{\wdchi}{\widetilde{\chi}}
\nc{\wdH}{\widetilde{H}}
\nc{\wdN}{\widetilde{N}}
\nc{\wdM}{\widetilde{M}}
\nc{\wdO}{\widetilde{O}}
\nc{\wdR}{\widetilde{R}}
\nc{\wdS}{\widetilde{S}}
\nc{\wdV}{\widetilde{V}}

\nc{\wdC}{\widetilde{C}}

\nc{\Obj}{\operatorname{Obj}}
\nc{\Dglie}{\operatorname{{\mathcal D}glie}}
\nc{\Fin}{\operatorname{{\mathcal F}in}}

\nc{\Adm}{\operatorname{\mathcal{A}dm}}
\nc{\Sg}{{\cS(\fg)}}
\nc{\Shg}{{\cS(\fhg)}}
\nc{\Ug}{{\cU(\fg)}}
\nc{\Uhg}{{\cU(\fhg)}}
\nc{\Sh}{{\cS(\fh)}}
\nc{\Uh}{{\cU(\fh)}}
\nc{\Uhh}{{\cU(\fhh)}}
\nc{\Zg}{{{\mathcal{Z}}(\fg)}}

\nc{\Vir}{{\mathcal{V}ir}}

\nc{\tZg}{{\widetilde{\mathcal Z}({\mathfrak g})}}
\nc{\Zk}{{\mathcal Z}({\mathfrak k})}

\nc{\Up}{{\mathcal U}({\mathfrak p})}
\nc{\Ah}{{\mathcal A}({\mathfrak h})}
\nc{\Ag}{{\mathcal A}({\mathfrak g})}
\nc{\Ap}{{\mathcal A}({\mathfrak p})}
\nc{\Zp}{{\mathcal Z}({\mathfrak p})}
\nc{\cZ}{\mathcal Z}
\nc{\cS}{\mathcal S}
\nc{\cT}{\mathcal{T}}

\nc{\cA}{\mathcal A}
\nc{\cU}{\mathcal U}
\nc{\cH}{\mathcal H}
\nc{\cM}{\mathcal M}
\nc{\cL}{\mathcal L}
\nc{\cF}{\mathcal F}
\nc{\fg}{\mathfrak g}

\nc{\fo}{\mathfrak o}

\nc{\CO}{\mathcal O}
\nc{\Cl}{\mathcal {C}\ell}

\nc{\cR}{\mathcal{R}}
\nc{\bM}{\mathbf{M}}
\nc{\bL}{\mathbf{L}}
\nc{\bN}{\mathbf{N}}

\nc{\zq}{\mathpzc q}

\nc{\fl}{\mathfrak l}
\nc{\fn}{\mathfrak n}
\nc{\fm}{\mathfrak m}
\nc{\fp}{\mathfrak p}
\nc{\fh}{\mathfrak h}
\nc{\ft}{\mathfrak t}
\nc{\fk}{\mathfrak k}
\nc{\fb}{\mathfrak b}
\nc{\fs}{\mathfrak s}
\nc{\fB}{\mathfrak B}

\nc{\vareps}{\varepsilon}
\nc{\varesp}{\varepsilon}
\nc{\veps}{\varepsilon}

\nc{\fsl}{\mathfrak{sl}}
\nc{\fpsl}{\mathfrak{psl}}
\nc{\fgl}{\mathfrak{gl}}
\nc{\fso}{\mathfrak{so}}

\nc{\fpq}{\mathfrak{pq}}
\nc{\fq}{\mathfrak q}
\nc{\fsq}{\mathfrak{sq}}
\nc{\fpsq}{\mathfrak{psq}}

\nc{\fhg}{\hat{\fg}}
\nc{\fhn}{\hat{\fn}}
\nc{\fhh}{\hat{\fh}}
\nc{\fhb}{\hat{\fb}}
\nc{\hrho}{\hat{\rho}}

\nc{\hsl}{\hat{\fsl}}
\nc{\fpo}{\mathfrak{po}}
\nc{\dirlim}{\underset{\rightarrow}{\lim}\,}
\nc{\nen}{\newenvironment}
\nc{\ol}{\overline}
\nc{\ul}{\underline}
\nc{\ra}{\rightarrow}
\nc{\lra}{\longrightarrow}
\nc{\Lra}{\Longrightarrow}

\nc{\Lla}{\Longleftarrow}

\nc{\Llra}{\Longleftrightarrow}

\nc{\thla}{\twoheadleftarrow}

\nc{\hra}{\hookrightarrow}

\nc{\iso}{\overset{\sim}{\lra}}

\nc{\ssubset}{\underset{\not=}{\subset}}

\nc{\vac}{|0\rangle}

\nc{\Thm}[1]{Theorem~\ref{#1}}
\nc{\Prop}[1]{Proposition~\ref{#1}}
\nc{\Lem}[1]{Lemma~\ref{#1}}
\nc{\Cor}[1]{Corollary~\ref{#1}}
\nc{\Conj}[1]{Conjecture~\ref{#1}}
\nc{\Claim}[1]{Claim~\ref{#1}}
\nc{\Defn}[1]{Definition~\ref{#1}}
\nc{\Exa}[1]{Example~\ref{#1}}
\nc{\Rem}[1]{Remark~\ref{#1}}
\nc{\Note}[1]{Note~\ref{#1}}
\nc{\Quest}[1]{Question~\ref{#1}}
\nc{\Hyp}[1]{Hypoth\`ese~\ref{#1}}
\nen{thm}[1]{\label{#1}{\bf Theorem.\ } \em}{}
\nen{prop}[1]{\label{#1}{\bf Proposition.\ } \em}{}
\nen{lem}[1]{\label{#1}{\bf Lemma.\ } \em}{}
\nen{cor}[1]{\label{#1}{\bf Corollary.\ } \em}{}
\nen{conj}[1]{\label{#1}{\bf Conjecture.\ } \em}{}

\nen{claim}[1]{\label{#1}{\bf Claim.\ } \em}{}

\nen{defn}[1]{\label{#1}{\bf Definition.\ } }{}
\nen{exa}[1]{\label{#1}{\bf Example.\ } }{}
\nen{rem}[1]{\label{#1}{\em Remark.\ } }{}
\nen{note}[1]{\label{#1}{\em Note.\ } }{}
\nen{exer}[1]{\label{#1}{\em Exercise.\ } }{}
\nen{sket}[1]{\label{#1}{\em Sketch of proof.\ } }{}
\nen{quest}[1]{\label{#1}{\bf Question.\ } \em}{}

\nen{hyp}[1]{\label{#1}{\bf Hypoth\`ese.\ } \em}{}
\begin{document}
\setcounter{section}{-1}

\title[Weyl denominator identity]{Weyl denominator identity
for finite-dimensional Lie superalgebras}
\author[Maria Gorelik]{Maria Gorelik}

\address{Dept. of Mathematics, The Weizmann Institute of Science,
Rehovot 76100, Israel}
\email{maria.gorelik@weizmann.ac.il}
\thanks{Supported in part by ISF Grant No. 1142/07}

\begin{abstract}
Weyl denominator identity for the basic simple Lie superalgebras 
was formulated by V.~Kac and M.~Wakimoto  
and was proven by them for the defect one case. In this paper
we prove the identity for the rest of the cases.
\end{abstract}

\maketitle

\section{Introduction}\label{intro}
The basic simple Lie  superalgebras are finite-dimensional simple Lie
superalgebras, which have a reductive even part and admit an 
even non-degenerate invariant bilinear form. These algebras 
 were classified by V.~Kac in~\cite{Ksuper} and the list (excluding Lie algebra
case) consists of four series: $A(m,n), B(m,n), C(m), D(m,n)$ 
and the exceptional algebras $D(2,1,a), F(4), G(3)$.

Let $\fg$ be a 
basic simple Lie  superalgebra with a fixed triangular decomposition
$\fg=\fn_-\oplus\fh\oplus\fn_+$,
and let $\Delta_+=\Delta_{+,0}\coprod \Delta_{+,1}$ be
the corresponding set of positive roots.
The Weyl denominator associated to the above data is
$$R:=\frac{\prod_{\alpha\in\Delta_{+,0}}(1-e^{-\alpha})}
{\prod_{\alpha\in\Delta_{+,1}}(1+e^{-\alpha})}.$$

If $\fg$ is a finite-dimensional simple Lie algebra
(i.e., $\Delta_1=\emptyset$), then the Weyl denominator is given by the
Weyl denominator identity 
$$Re^{\rho}=\sum_{w\in W} \sgn(w)e^{w\rho},$$
where $\rho$ is the half-sum of the positive roots, $W$ is the Weyl
group, i.e. the subgroup of $GL(\fh^*)$ generated by the reflections
with respect to the roots, and $\sgn(w)\in\{\pm 1\}$ is the sign of $w\in W$.
This identity may be viewed as the character
of trivial representation of the corresponding Lie algebra.

The Weyl denominator identities for superalgebras were formulated
and partially proven (for $A(m-1,n-1), B(m,n), D(m,n)$ with $\min(m,n)=1$
and for $C(n), D(2,1,a), F(4), G(3)$) by 
V.~Kac and M.~Wakimoto in~\cite{KW94}.
In order to state the Weyl denominator identity for  basic simple Lie 
superalgebras we need the following notation. 
Let $\Delta^{\#}$ be the ``largest'' 
component of $\Delta_0$, see~\ref{Delfin} for the definition.
Let $W$ be the Weyl group of $\fg_0$, i.e. the subgroup of 
$GL(\fh^*)$ generated by the reflections 
with respect to the even roots $\Delta_0$, 
and let $\sgn(w)$ be the sign of $w$. One has $W=W_1\times W_2$, where 
$W^{\#}$ be the Weyl group of root system $\Delta^{\#}$,
i.e. the subgroup of $W$ generated by the reflections
with respect to the roots from $\Delta^{\#}$. 
Set $\rho_0:=\sum_{\alpha\in\Delta_{+,0}}\alpha/2,\ \ 
\rho_1:=\sum_{\alpha\in\Delta_{+,1}}\alpha/2,\ \
\rho:=\rho_0-\rho_1$.
A subset $\Pi$ of $\Delta_+$ is called a {\em set of simple roots} if
the elements of $\Pi$ are linearly independent and 
$\Delta_+\subset \sum_{\alpha\in \Pi}\mathbb{Z}_{\geq 0}\alpha$.
For each functional $f$ the corresponding set $\Delta_+(f)$ contains
a unique system of simple roots, which we denote by $\Pi(f)$.
A subset $S$ of $\Delta$ is called {\em maximal isotropic} 
if the elements of $S$ form a basis of a maximal isotropic space in 
$V$.  By~\cite{KW94}, $\Delta$ contains a maximal isotropic subset
and each maximal isotropic subset  is a subset 
of a set of simple roots (for a certain functional $f$).
Fix a  maximal isotropic subset $S\subset \Delta$, choose
a set of simple roots $\Pi$ containing $S$, and choose a functional
$f:V\to \mathbb{R}$ in such a way that $\Pi=\Pi(f)$. Let $R$ be
the Weyl denominator for this choice of $f$.
The following Weyl denominator identity  was 
suggested by V.~Kac and M.~Wakimoto in~\cite{KW94}:
\begin{equation}\label{denomKW}
Re^{\rho}=\sum_{w\in W^{\#}} \sgn(w) 
w\bigl(\frac{e^{\rho}}{\prod_{\beta\in S}(1+e^{-\beta})}\bigr).
\end{equation}

If $S$ is empty (i.e., $(-,-)$ is a positive/negative definite)
the denominator identity takes the form 
$Re^{\rho}=\sum_{w\in W} \sgn(w)e^{w\rho}$. In this case
either $\Delta_1$ is empty (i.e., $\fg$ is a Lie algebra),
or $\fg=\mathfrak{osp}(1,2l)$ (type $B(0,l)$). 
The Weyl denominator identity  for the case $\mathfrak{osp}(1,2l)$ was proven
 in~\cite{K77}; for the case when
 $S$ has the cardinality one the identity was proven in~\cite{KW94}.

The Weyl denominator identity for root system of Lie (super)algebra $\fg$
can be again naturally interpreted as the character of 
one-dimensional representation of $\fg$.
The proofs in   abovementioned
 cases (\cite{K77},\cite{KW94}) are based on a analysis
of the highest weights of irreducible subquotients of the Verma module
$M(0)$ over $\fg$. In this paper we give a proof of the 
Weyl denominator identity~(\ref{denomKW}) for the case when
 $S$ has the cardinality greater than one.
A similar proof works for the case when the cardinality of $S$ is one. 
Unfortunately, our proof does not
use representation theory, but requires an analysis of the roots systems. 
The proof is based on a case-by-case verification of the following facts:

(i) the monomials appearing  in the right-hand side of~(\ref{denomKW}) 
are of  the form $e^{\rho-\nu}$, where $\nu\in Q^+:=\sum_{\alpha\in\Pi}
\mathbb{Z}_{\geq 0}\alpha$ and that the coefficient of $e^{\rho}$ is one;

(ii) the right-hand side of~(\ref{denomKW}) is $W$-skew-invariant
(i.e., $w\in W$ acts by the multiplication by $\sgn(w)$).
 
Taking into account that for any $\lambda\in V$ the stabilizer of $\lambda$
in $W$ is either trivial or contains a reflection, and that if
the stabilizer is trivial and
$W\lambda\subset (\rho_0-\sum_{\alpha\in\Delta_+}\mathbb{Z}_{\geq 0}\alpha)$,
then $\lambda=\rho_0$, we easily deduce the identity~(\ref{denomKW}) from 
(i), (ii).

I.~M.~Musson informed us that he has an unpublished proof of  
the Weyl denominator identity for basic simple Lie superalgebras.

The Weyl denominator identity for the affinization of a simple 
finite-dimensional 
Lie superalgebra with non-zero Killing form was also formulated by
V.~Kac and M.~Wakimoto in~\cite{KW94} and was proven for the defect one case.
We prove this identity in~\cite{G}.

{\em Acknowledgments.} 
I am very grateful to V.~Kac for his  patience and useful comments.
I would like to thank  D.~Novikov and A.~Novikov for their support.

\section{The algebra $\cR$}
In this section we introduce the algebra $\cR$. Since
the main technical difficulty in our proof of the denominator
identity comes from the existence
of different triangular decompositions, we illustrate
how the proof works when there is only one  triangular decomposition
(see~\ref{liecase},\ref{qn}).

\subsection{Notation}\label{notat}
Denote by $Q$ the root lattice of $\fg$ and by $Q^+$
the positive part of root lattice $Q^+:=
\sum_{\alpha\in\Pi} \mathbb{Z}_{\geq 0}\alpha$;
we introduce the following partial order on $V$:
$$\mu\leq\nu\ \text{ if } (\nu-\mu)\in\sum_{\alpha\in\Delta_+}
\mathbb{R}_{\geq 0}\alpha.$$
Introduce the height function $\htt:Q^+\to\mathbb{Z}_{\geq 0}$ by 
$$\ \htt
(\sum_{\alpha\in\Pi}m_{\alpha}\alpha):=\sum_{\alpha\in\Pi}m_{\alpha}.$$

We use the following notation: for $X\subset \mathbb{R}, Y\subset V$
we set $XY:=\{xy|x\in X, y\in Y\}$; for instance,
$Q^+:= \mathbb{Z}_{\geq 0}\Pi$.

Note that  $\Delta_0$ is the root system of
the reductive Lie algebra $\fg_0$.  In particular, all isotropic roots
are odd. Both sets $\Delta_0, \Delta_1$ are $W$-stable:
$W\Delta_i=\Delta_i$.

\subsection{The algebra $\cR$}\label{supp}
Denote by $\mathbb{Q}[e^{\nu},\ \nu\in V]$ the algebra of polynomials
in $e^{\nu},\nu\in V$.
Let $\cR$ be the algebra of rational functions of the form
$$\frac{X}{\prod_{\alpha\in\Delta_+}
(1+a_{\alpha}e^{-\alpha})^{m_{\alpha}}},$$
where $X\in \mathbb{Q}[e^{\nu},\ \nu\in V]$
and $a_{\alpha}\in \mathbb{Q}$, 
$m_{\alpha}\in\mathbb{Z}_{\geq 0}$. Clearly, $\cR$ contains
the rational functions of the form
$\frac{X}{\prod_{\alpha\in\Delta}(1+a_{\alpha}e^{-\alpha})^{m_{\alpha}}}$,
where $X,a_{\alpha},m_{\alpha}$ as above.
The group $W$ acts on $\cR$ by the automorphism mapping $e^{\nu}$
to $e^{w\nu}$. Say that $P\in\cR$ is {\em $W$-invariant}
(resp., {\em $W$-skew-invariant}) if  $wP=P$ (resp.,
$wP=\sgn(w)P$) for every $w\in W$.

\subsubsection{}
For a sum $Y:=\sum b_{\mu} e^{\mu}, b_{\mu}\in\mathbb{Q}$
introduce the {\em support } of $Y$ by the formula
$$\supp(Y):=\{\mu|\ b_{\mu}\not=0\}.$$
Any element of $\cR$ can be uniquely expanded in the form
$$\frac{\sum_{i=1}^m a_ie^{\nu_i}}{\prod_{\alpha\in\Delta_+}
(1+a_{\alpha}e^{-\alpha})^{m_{\alpha}}}=\sum_{i=1}^s
\sum_{\mu\in Q^+} b_{\mu}e^{\nu_i-\mu},\ b_{\mu}\in\mathbb{Q}.$$
For $Y\in\cR$ denote by $\supp(Y)$ the support of its expansion;
by above, $\supp(Y)$ lies in a finite union of cones 
of the form $\nu-Q^+$.

\subsubsection{}\label{regorb}
We call $\lambda\in V$ 
{\em regular} if $\Stab_W \lambda=\{\id\}$; we 
call the orbit $W\lambda$  regular if $\lambda$ is regular
(so the orbit consists of regular points).

It is well-known that  for $\lambda\in V$ the stabilizer
$\Stab_W \lambda$ is either trivial or contains a reflection 
(see~\Lem{cor1} (ii)).
Therefore the stabilizer of a non-regular point $\lambda\in V$
contains a reflection. As a result, the space of $W$-skew-invariant
elements of $\mathbb{Q}[e^{\nu},\ \nu\in V]$ 
is spanned by $\sum_{w\in W} \sgn(w) e^{w\lambda}$, where
$\lambda\in V$ is regular. In particular, the support of
a $W$-skew-invariant element of $\mathbb{Q}[e^{\nu},\ \nu\in V]$ 
is a union of regular $W$-orbits.

\subsection{Lie algebra case}\label{liecase}
The denominator identity for Lie algebra is
$$Re^{\rho}=\prod_{\alpha\in\Delta_+} (1-e^{-\alpha})e^{\rho}
=\sum_{w\in W}\sgn(w) e^{w\rho}.$$

There are several proofs of this identity. The proof which 
we are going to generalize is the following.
Observe that $Re^{\rho}\in\mathbb{Q}[e^{\nu},\ \nu\in V]$ and 
$\supp(Re^{\rho})\subset (\rho-Q^+)$. Moreover, 
$Re^{\rho}=\prod_{\alpha\in\Delta_+} (e^{\alpha/2}-e^{-\alpha/2})$
is $W$-skew-invariant,
so $\supp(Re^{\rho})$ is a union of regular
orbits lying in $(\rho-Q^+)$. However, $W\rho$ is
the only regular orbit lying entirely in $(\rho-Q^+)$, see~\Lem{cor1} (iii). 
Hence $Re^{\rho}$ is proportional to $\sum_{w\in W}\sgn(w) e^{w\rho}$.
Since the coefficient of $e^{\rho}$ in the expression $Re^{\rho}$ is $1$,
the coefficient of proportionality is $1$.

\subsection{Case $Q(n)$}\label{qn}
For the case $\fg:=Q(n)$ one has $\fg_0=\fgl(n)$, $W=S_n$,
$\Delta_{0+}=\Delta_{+,1}=\{\vareps_i-\vareps_j\}_{1\leq i<j\leq n}$ and 
so $\rho_0=\rho_1,\ \rho=0$.
The Weyl denominator is $R=\prod_{\alpha\in\Delta_{+,0}}\frac{1-e^{-\alpha}}
{1+e^{-\alpha}}$. For each  $S\subset\Delta_{+,1}$ define
$$A(S):=\{w\in S_n|\ wS\subset\Delta_{+,0}\},
\ \ a(S):=\sum_{w\in A(S)} \sgn(w).$$

\subsubsection{}
\begin{prop}{}
For each  $S\subset\Delta_{+,1}$ one has
$$a(S)R=\sum_{w\in S_n} 
\sgn(w)\frac{1}{\prod_{\beta\in S} (1+e^{-w\beta})}.$$
\end{prop}
\begin{proof}
Observe that the support of both sides of the formula
lies in $-Q^+$ and that the coefficients of $1=e^0$ in both sides
are equal to $a(S)$.
Multiplying both sides of the formula by the $W$-invariant expression
$\prod_{\alpha\in\Delta_{+,1}}(1+e^{-\alpha})e^{\rho_1}=
\prod_{\alpha\in\Delta_{+,1}}(e^{\alpha/2}+e^{-\alpha/2})$ we obtain
$$\begin{array}{ll}
Y:&=\prod_{\alpha\in\Delta_{+,0}}(1-e^{-\alpha})e^{\rho_0}-
\prod_{\alpha\in\Delta_{+,1}}(1+e^{-\alpha})e^{\rho_1}
\sum_{w\in S_n} \sgn(w)\frac{1}{\prod_{\beta\in S} (1+e^{-w\beta})}\\
&=\prod_{\alpha\in\Delta_{+,0}}(1-e^{-\alpha})e^{\rho_0}-
\sum_{w\in S_n} \sgn(w)w\bigl(\prod_{\alpha\in\Delta_{+,0}\setminus S}
(1+e^{-\alpha})e^{\rho_0}\bigr).
\end{array}$$
Since $\prod_{\alpha\in\Delta_{+,0}}(1-e^{-\alpha})e^{\rho_0}=
\prod_{\alpha\in\Delta_{+,0}}(e^{\alpha/2}-e^{-\alpha/2})$
is $W$-skew-invariant, $Y$ is also $W$-skew-invariant. Clearly, 
$Y\in\mathbb{Q}[e^{\nu},\ \nu\in V]$, so, by~\ref{regorb},
$\supp(Y)$ is a union of regular $W$-orbits.
By above, $\supp(Y)\subset (\rho_0-Q^+)\setminus\{\rho_0\}$.
However, by~\Lem{cor1} (iii), any regular $W$-orbit intersects
$\rho_0+Q^+$. Hence $Y=0$ as required.
\end{proof}

\subsubsection{}
In order to obtain a formula for the Weyl denominator $R$, 
we choose $S$ such that $a(S)\not=0$. Taking
$S=\{\vareps_1-\vareps_n,\vareps_2-\vareps_{n-1},\ldots,
\vareps_{[\frac{n}{2}]}-\vareps_{n+1-[\frac{n}{2}]}\}$
and using~\Lem{lemQ}, we obtain the following formula
$$R=\frac{1}{[n/2]!}\sum_{w\in S_n} 
\sgn(w)\frac{1}{\prod_{\beta\in S} (1+e^{-w\beta})},$$
which appears in~\cite{KW94} (7.1) (up to a constant factor).

Note that such $S$ has a minimal cardinality:
if the cardinality of $S$ is less than
$[\frac{n}{2}]$, then $a(S)=0$. Indeed, if the cardinality of $S$ is less than
$[\frac{n}{2}]$, then there is
a root $\vareps_i-\vareps_j$, which does not belong to the span of $S$,
 and thus $s_{\vareps_i-\vareps_j}W(S)=W(S)$; since 
$\sgn(w)+\sgn(s_{\vareps_i-\vareps_j}w)=0$, this forces $a(S)=0$.

\subsubsection{}
\begin{lem}{lemQ}
Set
$A:=\{\sigma\in S_n|\ \forall i\leq \frac{n}{2}\ \ \
\sigma(i)>\sigma(n+1-i)\}$.
Then
$$\sum_{\sigma\in A} \sgn(\sigma)=[n/2]!$$
\end{lem}
\begin{proof}
For each $\sigma\in A$ let $P(\sigma)$ be the set of pairs
$\{(\sigma(1),\sigma(n);(\sigma(2),\sigma(n-1));\ldots\}$,
that is $P(\sigma):=\{(\sigma(j),\sigma(n+1-j))\}_{j=1}^{[\frac{n}{2}]}$.
Let $B:=\{\sigma\in A|\  P(\sigma)=
\{(j,j+1)\}_{j=1}^{[\frac{n}{2}]}\}$. 
Define an involution $f$ on the set $A\setminus B$ 
as follows: for $\sigma\in A\setminus B$ set $f(\sigma):=(i,i+1)\circ\sigma$,
where $i$ is minimal such $(i,i+1)\not\in  P(\sigma)$. Since 
$\sgn(\sigma)+\sgn(f(\sigma))=0$, we get $\sum_{\sigma\in A\setminus B} 
\sgn(\sigma)=0$.
One readily sees  that $B$ has $[n/2]!$ elements and that
$\sgn(\sigma)=1$ for each $\sigma\in B$. Hence 
$\sum_{\sigma\in A} \sgn(\sigma)=[n/2]!$ as required.
\end{proof}

\section{Notation}
Let $\fg=\fg_0\oplus\fg_1$ be a a basic Lie superalgebra
with a fixed triangular decomposition of the even part:
$\fg_0=\fn_{-,0}\oplus\fh\oplus\fn_{+,0}$. 
For $A(m-1,n-1)$-type we put  $\fg=\fgl(m|n)$ 
(one readily sees that the denominator identities for $\fgl(m|n)$ imply
the denominator identities for $\fsl(m|n), \fpsl(n|n)$). Let
$\Delta_0$ (resp., $\Delta_1$) be the set of even (resp., odd) roots
of $\fg$. Set $V=\fh^*_{\mathbb{R}}$ (so $V=\spn\Delta$ for
$\fg\not=A(m,n)$ and $V=\mathbb{R}\spn\Delta\oplus\mathbb{R}$
for $\fg=A(m,n)$). Denote by $(-,-)$  a non-degenerate
symmetric bilinear form on $V$, induced by a non-degenerate
invariant bilinear form on $\fg$. Retain notation of Section~\ref{intro}
and  define the Weyl denominator.

The dimension of a maximal isotropic space in 
$V=\fh^*_{\mathbb{R}}$ is called the {\em
defect} of $\fg$.
If $\fg$ is a Lie algebra or $\fg=\mathfrak{osp}(1,2l)$ (type $B(0,l)$)
then the defect of $\fg$ is zero;
the defect of $A(m-1,n-1), B(m,n), D(m,n)$ is equal to $\min(m,n)$;
for $C(n)$ and the exceptional Lie superalgebras the defect is equal to one.
Notice that the cardinality of a maximal isotropic set $S$ is equal to
defect of $\fg$.

\subsection{Admissible pairs}\label{adm}
The set of positive even roots  $\Delta_{+,0}$
is determined by the triangular decomposition 
$\fg_0=\fn_{-,0}\oplus\fh\oplus\fn_{+,0}$
(i.e., $\Delta_{+,0}$ is the set of weights of $\fn_{+,0}$).
Recall that  for each  maximal isotropic set $S$ there exists
a set of simple roots containing $S$.
We call a pair $(S,\Pi)$ {\em admissible} if 
$S\subset\Pi$ is a maximal isotropic set of roots
and $\Pi$ is a set of simple roots
such that the corresponding set of positive even roots coincides with
  $\Delta_{+,0}$:
$$(S,\Pi)\text{ is admissible if }
S\subset\Pi\ \&\  \Delta_+(\Pi)\cap\Delta_0=\Delta_{+,0}.$$

For a fixed set of simple roots $\Pi$ we retain notation of~\ref{notat} 
and~\ref{supp}.

\subsection{The set $\Delta^{\#}$}\label{Delfin}
Let $\Delta_1,\Delta_2$ be two finite irreducible root systems;
we say that $\Delta_1$ is ``larger'' than $\Delta_2$ if
either the rank of $\Delta_1$ is greater than the rank of $\Delta_2$, or
the ranks are equal and $\Delta_1\subset\Delta_2$.

If the defect of $\fg$ is greater than one, then the root
system $\Delta_0$ is a disjoint union of two irreducible root systems.
We denote by $\Delta^{\#}$ the irreducible component, 
which is not the smallest
one, i.e. $\Delta_0=\Delta^{\#}\coprod\Delta_2$, where 
$\Delta^{\#}$ is not smaller, than $\Delta_2$, see the following table:

$$\begin{tabular}{|l || l | l| l|}
\hline
$\Delta$ & \ \ \ A(m-1,n-1) & \ \ \ \ \ \ \ \ \  B(m,n) & \ \ \ D(m,n)\\
 & $m>n$\ \  \ \ $m\leq n$ & $m>n$\ \  \ $m<n$\ \ \ \ \ $m=n$              
 & $m>n$\ \ \ $m\leq n$\\
$\Delta^{\#}$ & $A_{m-1}$\ \  \ \ \ \ $A_{n-1}$   & \ $B_m$\ \ \ \ \ \ $C_n$
\ \  \ \ \ \ \ \ \ $B_m$ or $C_m$ & $D_m$\  \ \ \  \ \ \ \ \ $C_n$\\
\hline
\end{tabular}$$

The notion of $\Delta^{\#}$ in~\cite{KW94} coincides 
with the above one, except for the case $B(m,m)$, where we allow
both choices $B_m$ and $C_m$, whereas in~\cite{KW94} 
$\Delta^{\#}$ is of the type $C_m$.

Notice that $\fg_0=\fs_1\times \fs_2$, where $\fs_1,\fs_2$ are 
reductive Lie algebras and 
 $\Delta^{\#},\Delta_0\setminus\Delta^{\#}$ are roots systems
of $\fs_1, \fs_2$ respectively.
We normalize $(-,-)$ in such a way that 
$\Delta^{\#}:=\{\alpha\in\Delta_0|\ (\alpha,\alpha)>0\}$; then
$\Delta_0\setminus\Delta^{\#}=\{\alpha\in\Delta_0|\ (\alpha,\alpha)<0\}$.

\section{Outline of the proof}
\subsection{}\label{X}
Let $\fg$ be  one of the Lie superalgebras $A(m-1,n-1), B(m,n), D(m,n),
m,n>0$.

\subsection{Expansion of the right-hand side of~(1)}
Let $(S,\Pi)$ be an admissible pair. Set
\begin{equation}\label{defX}
X:=\sum_{w\in W^{\#}}\sgn(w)w\bigl(\frac{e^{\rho}}{\prod_{\beta\in S}
(1+e^{-\beta})}\bigr)
\end{equation}
and rewrite the denominator identity~(\ref{denomKW}) as $Re^{\rho}=X$.

Expanding $X$ we obtain
\begin{equation}\label{expX}
\begin{array}{l}
X=\sum_{w\in W^{\#}}\sum_{\mu\in\mathbb{Z}_{\geq 0}S} \sgn(w)(-1)^{\htt \mu}
e^{\varphi(w)-|w|\mu+w\rho},
\end{array}
\end{equation}
where
$$\varphi(w):=\!\!\!\!\sum_{\beta\in S: w\beta<0}\!\! \!\!w\beta\in -Q^+$$
and  $|w|$ is a linear map $\mathbb{Z}_{\geq 0}S\to Q^+$ defined
on $\beta\in S$ by the formula
$$|w|\beta=\left\{ \begin{array}{ll}w\beta & \text{ for }
w\beta>0,\\
-w\beta & \text{ for } w\beta<0.\end{array}\right.$$

\subsection{Main steps}
The proof has the following steps:

(i) We introduce certain operations on the admissible pairs $(S,\Pi)$ and show
that these operations preserve  the expressions
$X$ and $Re^{\rho}$. Consider the equivalence
relation on the set of admissible pairs $(S,\Pi)$
generated by these operations. We will show that there are
two equivalence classes for $D(m,n), m>n$ 
and one equivalence class for other cases.

(ii) We check that $\supp(X)\subset (\rho-Q^+)$ and that
the coefficient of $e^{\rho}$ in $X$ is $1$ for a certain choice
of $(S,\Pi)$ (for $D(m,n), m>n$ we check this
for $(S,\Pi)$ and $(S',\Pi)$, which
are representatives of the equivalence classes).

(iii) We show that $X$ is $W$-skew-invariant for a certain choice
of $(S,\Pi)$ (for $D(m,n), m>n$ we show this
for $(S,\Pi)$ and $(S',\Pi)$, which
are representatives of the equivalence classes).

For $\fg$ of $A(n-1,n-1)$ type we change (ii) to (ii'):

(ii') We check, for a certain choice of $(S,\Pi)$,
that $\supp(X)\subset (\rho-Q^+)$ and that for $\xi:=\sum_{\beta\in S}\beta$
the coefficients of $e^{\rho-s\xi}$ in $X$ and in $Re^{\rho}$ 
are equal for each $s\in\mathbb{Z}_{\geq 0}$.

The choices of $(S,\Pi)$ in (ii), (iii) are the same only for $A(m,n)$ case.

\subsection{Why (i)--(iii) imply~(1)}\label{stepsfin}
Let us show that (i)--(iii) imply the denominator identity $X=Re^{\rho}$.
Indeed, assume that $X-Re^{\rho}\not=0$.

Since $WS\subset \Delta_{1}$, $X-Re^{\rho}$ is a rational function 
with the denominator of the form
$\prod_{\beta\in \Delta_{1}^+}(1+e^{-\beta})$; we write
$$X-Re^{\rho}=\frac{Y}
{\prod_{\beta\in \Delta_{1}^+}(1+e^{-\beta})}=
\frac{Ye^{\rho_1}}
{\prod_{\beta\in \Delta_1^+}(e^{\beta/2}+e^{-\beta/2})},$$
where $Y\in\mathbb{Q}[e^{\nu},\nu\in V]$. One has
$$Re^{\rho}=\frac{\prod_{\alpha\in\Delta_0^+} (e^{\alpha/2}-e^{-\alpha/2})}
{\prod_{\alpha\in\Delta_1^+} (e^{\alpha/2}+e^{-\alpha/2})}$$
and the latter expression is $W$-skew-invariant, since its
enumerator  is $W$-skew-invariant and its denominator is $W$-invariant. 
Combining (i) and (iii), we obtain that
 $X-Re^{\rho}$ is $W$-skew-invariant. Thus $Ye^{\rho_1}$ is 
a $W$-skew-invariant element of $\mathbb{Q}[e^{\nu},\nu\in V]$ and
so $\supp(Ye^{\rho_1})$ is a union of regular orbits.

Observe that $\supp (Re^{\rho})\subset (\rho-Q^+)$ and that
the coefficient of $e^{\rho}$  in $Re^{\rho}$ is $1$. Using (i), (ii) we get
$\supp (X-Re^{\rho})\subset  (\rho-Q^+)\setminus \{\rho\}$.
Note that the sets of maximal elements in $\supp (Y)$
and in $\supp (X-Re^{\rho})$ coincide. Thus
$\supp Y\subset  (\rho-Q^+)\setminus \{\rho\}$  that is
$\supp (Ye^{\rho_1})\subset (\rho_0-Q^+)\setminus \{\rho_0\}$.
Hence $\supp(Ye^{\rho_1})$ is a union of
regular orbits lying in $(\rho_0-Q^+)\setminus \{\rho_0\}$.

By~\ref{regorbit}, for $\fg\not=\fgl(n|n)$,
the set $(\rho_0-Q^+)\setminus \{\rho_0\}$ does not contain
 regular $W$-orbit, a contradiction.

Let $\fg=\fgl(n|n)$. Choose $\Pi$ as in~\ref{rootsys}.
By~\ref{regorbit} the regular orbits in $(\rho_0-Q^+)$ 
are of the form $W(\rho_0-s\xi)$
with $s\in\mathbb{Z}_{\geq 0},\ 
\xi=\sum_{\beta\in S}\beta$. One has $W\xi=\xi$ and $w\rho_0\leq \rho_0$,
so $\rho_0-s\xi$ is the maximal element in its $W$-orbit.
Thus a maximal element in $\supp Ye^{\rho_1}$ is of the form 
$\rho_0-s\xi$, so a maximal element in $\supp Y$ is $\rho-s\xi$.
Then,  by above, $\rho-s\xi\in \supp (X-Re^{\rho})$, 
which contradicts to (ii').

\section{Regular orbits}
\subsection{}
Let $\fg$ be a reductive finite-dimensional   Lie algebra, let $W$ be its
Weyl group, let $\Pi$ be its set of simple roots 
and let $\Pi^{\vee}$ be the set of simple 
coroots. For $\rho$, defined as above, one has
$\langle \rho,\alpha^{\vee}\rangle=1$ for each $\alpha\in\Pi$.
Set
$$Q_{\mathbb{Q}}=\sum_{\alpha\in\Pi} \mathbb{Q}\alpha,\ \  
Q_{\mathbb{Q}}^+:=\sum_{\alpha\in\Pi} \mathbb{Q}_{\geq 0}\alpha.$$

As above, we define  partial order on $\fh^*_{\mathbb{R}}$ by the formula
$\mu\leq\nu\ \text{ if } (\nu-\mu)\in\sum_{\alpha\in\Delta_+}
\mathbb{R}_{\geq 0}\alpha$.

Let $P\subset \fh^*_{\mathbb{R}}$ be the weight lattice  of $\fg$, 
i.e. $\nu\in P$ iff $\langle \nu,\alpha^{\vee}\rangle\in\mathbb{Z}$ 
for any $\alpha\in\Pi$, and let $P^+$ be the positive part of $P$, i.e.
$\nu\in P^+$ iff $\langle \nu,\alpha^{\vee}\rangle\in\mathbb{Z}_{\geq 0}$ 
for any $\alpha\in\Pi$. One has $P\subset Q_{\mathbb{Q}}$.

\subsubsection{}
\begin{lem}{cor1}
(i) $P=WP^+$.

(ii) For any $\lambda\in\fh^*_{\mathbb{R}}$ the stabilizer of $\lambda$
in $W$ is either trivial or contains a reflection.

(iii) A regular orbit in $P$ intersects with the set $\rho+Q^+_{\mathbb{Q}}$.
\end{lem}
\begin{proof}
The group $W$ is finite. Take $\lambda\in\fh^*_{\mathbb{R}}$ and let
$\lambda'=w\lambda$ be a maximal element in the orbit $W\lambda$. Since
 $\lambda'$ is maximal, $\langle \lambda',\alpha^{\vee}\rangle\geq 0$ 
for each $\alpha\in\Pi$. 

For $\lambda\in P$ one has $\lambda'\in P$, so
$\langle \lambda',\alpha^{\vee}\rangle\in\mathbb{Z}_{\geq 0}$ 
for each $\alpha\in\Pi$, that is $\lambda'\in P^+$, hence (i).

For (ii) note that, if $\langle \lambda',\alpha^{\vee}\rangle=0$ 
for some $\alpha\in\Pi$, then $s_{\alpha}\in \Stab_{W}\lambda'$,
so $s_{w^{-1}\alpha}\in \Stab_{W}\lambda$. Assume that
$\langle \lambda',\alpha^{\vee}\rangle>0$  for all $\alpha\in\Pi$.
Take $y\in W, y\not=\id$ and write $y=y's_{\alpha}$ for $\alpha\in\Pi$, where
the length of $y'\in W$ is less than the length of $y$. 
Then $y'\alpha\in\Delta_+$ (see, for instance,~\cite{Jbook}, A.1.1). 
One has $y'\lambda'-y's_{\alpha}\lambda'=
\langle \lambda',\alpha^{\vee}\rangle (y'\alpha)>0$ so  
$y'\lambda'>y's_{\alpha}\lambda'$. Now (ii) follows by the induction on the 
length of $y$.

For (iii) assume that $\lambda'$ is regular, that is 
$\langle \lambda',\alpha^{\vee}\rangle>0$  for all $\alpha\in\Pi$.
Since $\lambda'\in P$, one has
$\langle \lambda',\alpha^{\vee}\rangle\in\mathbb{Z}_{\geq 1}$ so
$\langle \lambda-\rho,\alpha^{\vee}\rangle\geq 0$ for all $\alpha\in\Pi$.
Write 
$\lambda-\rho=\sum_{\beta\in\Pi} x_{\beta}\beta$. For a vector
$(y_{\beta})_{\beta\in \Pi}$ write $y\geq 0$ if $y_{\beta}\geq 0$
for each $\beta$.
The condition $\langle \lambda-\rho,\alpha^{\vee}\rangle\geq 0$ 
for all $\alpha\in\Pi$ means that $Ax\geq 0$,
where $x=(x_{\beta})_{\beta\in\Pi}$
and $A=(\langle\alpha^{\vee},\beta\rangle)_{\alpha,\beta\in \Pi}$ 
is the Cartan matrix of $\fg$.
From~\cite{Kbook}, 4.3 it follows that  $Ax\geq 0$ forces
$x\geq 0$. Hence $\lambda-\rho\in Q^+_{\mathbb{Q}}$ as required.
\end{proof}

\subsection{}\label{regorbit}
Now let $\fg$ be a basic simple Lie superalgebra, $Q$ be 
its root lattice and $Q^+$ be the positive part of $Q$.

\subsubsection{}
\begin{cor}{cor2}
Let $\fg$  be  a basic simple Lie superalgebra
and $\fg\not=C(n), A(m,n)$.
A regular orbit in the root lattice $Q$
 intersects with the set $\rho_0+\mathbb{Q}_{\geq 0}\Delta_{+,0}$.
\end{cor}
\begin{proof}
For $\fg\not=C(n), A(m,n)$ one has $\mathbb{Q}\Delta_0=\mathbb{Q}\Delta$ 
and the $\fg$-root lattice $Q$ is a subset of weight lattice of $\fg_0$. 
Thus the assertion follows from~\Cor{cor1}.
\end{proof}

\subsubsection{Case $\fg=\fgl(m|n)$}
For $\fgl(m|n)$ one has
$$\mathbb{Q}\Delta=\mathbb{Q}\Delta_{0}\oplus \mathbb{Q}\xi,\ \ 
\text{ where }\ \xi:=\sum\vareps_i-\frac{m}{n}\sum\delta_j.$$

Choose a set of simple roots as in~\ref{rootsys}.

\begin{lem}{}
Let $\fg=\fgl(m|n)$.

(i) For $m\not=n$, $W\rho_0$ is the only
regular orbit lying entirely in $(\rho_0-Q^+)$.

(ii) For $m=n$ the regular orbits lying entirely in $(\rho_0-Q^+)$
are of the form $W(\rho_0-s\xi), s\geq 0$, where
$\xi =\sum\vareps_i-\sum\delta_j$.
\end{lem}
\begin{proof}
Let $\iota: \mathbb{Q}\Delta\to\mathbb{Q}\Delta_{0}$ 
be the projection along $\xi$ (i.e., $\Ker\iota=\mathbb{Q}\xi$).
Since $\xi$ is $W$-invaraint, $w\iota(\lambda)=\iota(w\lambda)$.
Let $W\lambda\subset(\rho_0-Q^+)$ be a regular orbit. Since
$\iota(Q)$ lies in the weight lattice of $\fg_0$,
$\iota(W\lambda)$ intersects with $\rho_0+\mathbb{Q}_{\geq 0}\Delta_{+,0}$,
by~\Cor{cor1}. Thus 
$W\lambda$ contains a point of the form $\rho_0+\nu+q\xi$,
where $\nu\in\mathbb{Q}_{\geq 0}\Delta_{+,0}$
and $q\in\mathbb{Q}$. By above, $\nu+q\xi\in -Q^+$
so $q\xi\in -\mathbb{Q}_{\geq 0}\Delta_+$.

Consider the case $m>n$. In this case, for any 
$\mu\in\mathbb{Q}_{\geq 0}\Delta_+$ one has
$(\mu,\vareps_1)\cdot (\mu,\vareps_m)\leq 0$.
Since $(\xi,\vareps_1)=(\xi,\vareps_m)\not=0$, the inclusion
$q\xi\in -\mathbb{Q}_{\geq 0}\Delta_+$ implies $q=0$.
Then $\nu\in\mathbb{Q}_{\geq 0}\Delta_{+,0}$ and $\nu=\nu+q\xi\in -Q^+$
so $\nu=0$. Therefore $W\lambda=W\rho_0$. Hence $W\rho_0$ is the only 
regular orbit lying entirely in $\rho_0-Q^+$.
Since $\fgl(m|n)\cong\fgl(n|m)$, this establishes (i).

Consider the case $m=n$. Set $\beta_i:=\vareps_i-\delta_i,
\beta'_i:=\delta_i-\vareps_{i+1}$.  One has $\xi=\sum_{i=1}^n\beta_i$
and $\Pi=\{\beta_i\}_{i=1}^n\cup\{\beta'_i\}_{i=1}^{n-1}$. The simple
roots of $\Delta_{+,0}$ are $\vareps_i-\vareps_{i+1}=
\beta_i+\beta'_i$ and $\delta_i-\delta_{i+1}=\beta'_i+\beta_{i+1}$. Thus
$\nu\in\mathbb{Q}_{\geq 0}\Delta_0^+$ takes the form
$\nu=\sum_{i=1}^{n-1}b_i(\beta_i+\beta_i')+c_i(\beta'_i+\beta_{i+1})$ 
with $b_i,c_i\in\mathbb{Q}_{\geq 0}$. By above, $\nu':=-(\nu+q\xi)\in Q^+$.
Therefore $\nu+\nu'\in\mathbb{Q}\sum_{i=1}^n\beta_i$ and
$\nu'\in Q^+$. One readily sees that this implies $b_i=c_i=0$
that is  $\nu=0$. One has $Q^+\cap \mathbb{Q}\xi=
\mathbb{Z}_{\geq 0}\xi$. Hence 
a regular orbit in $\rho_0-Q^+$ intersects with the set
$\rho_0-\mathbb{Z}_{\geq 0}\xi$ as required.
\end{proof}

\subsubsection{Case $C(n)$}\label{orbCn}
Take $\Pi=\{\vareps_1-\vareps_2,\vareps_2-\vareps_3,\ldots,
\vareps_n-\delta_1,\vareps_n+\delta_1\}$. One has
$\mathbb{Q}\Delta=\mathbb{Q}\Delta_{0}\oplus \mathbb{Q}\delta_1$.
We claim  that $W\rho_0$ is the only regular orbit
lying entirely in $(\rho_0-Q^+)$.

Indeed, take a regular orbit lying entirely in $(\rho_0-Q^+)$. 
Combining the fact that $W\delta_1=\delta_1$ and~\Cor{cor1}, we see that 
this orbit contains a point of the form $\rho_0+\nu+q\delta_1$,
where $\nu\in\mathbb{Q}_{\geq 0}\Delta_0^+$
and $q\in\mathbb{Q}$. Since $-(\nu+q\delta_1)\in Q^+$, one has
$q\delta_1\in -\mathbb{Q}_{\geq 0}Q^+$. However, $2\delta_1$ is
the difference of two simple roots
($2\delta_1=(\vareps_n+\delta_1)-(\vareps_n-\delta_1)$) so
$\mathbb{Q}\delta_1\cap (-\mathbb{Q}_{\geq 0}Q^+)=\{0\}$ that is
$q=0$. The conditions $\nu\in\mathbb{Q}_{\geq 0}\Delta_0^+$,
$-(\nu+q\delta_1)\in Q^+$ give $\nu=0$. The claim follows.

\section{Step (i)}
\label{sect(i)}
Consider the following operations with the admissible pairs $(S,\Pi)$.
 First type operations are the odd reflections 
$(S,\Pi)\mapsto (s_{\beta}S,s_{\beta}\Pi)=
(S\setminus\{\beta\}\cup\{-\beta\},s_{\beta}\Pi)$
with respect to an element of $\beta\in S$ (see~\ref{odd}).
By~\ref{odd}, these odd reflections preserve the expressions $X, Re^{\rho}$.
Second type operations are the operations
$(S,\Pi)\mapsto (S',\Pi)$ described in~\Lem{lemnewS}, where
it is shown that these operations also  preserve 
the expressions $X, Re^{\rho}$. Consider the equivalence
relation on the set of admissible pairs $(S,\Pi)$ 
generated by these operations. In~\ref{fPi} we will show that there are
two equivalence classes for $D(m,n), m>n$ 
and one equivalence class for other cases. In~\ref{fPi2}
we will  show that if $(S,\Pi), (S,\Pi')$ are admissible pairs, then 
$\Pi=\Pi'$ (on the other hand, there are admissible pairs
$(S,\Pi), (S',\Pi)$ with $S\not=S'$, see~\Lem{lemnewS}). 

\subsection{Notation}
Let us introduce the following operator $F:\cR\to\cR$
$$F(Y):=\sum_{w\in W^{\#}}\sgn(w) wY.$$
Clearly, $F(wY)=\sgn(w) F(Y)$ for $w\in W^{\#}$, so $F(Y)=0$ if $wY=Y$
for some $w\in W^{\#}$ with $\sgn(w)=-1$.

For an admissible pair $(S,\Pi)$ introduce
$$Y(S, \Pi):=\frac{e^{\rho(\Pi)}}{\prod_{\beta\in S}(1+e^{-\beta})},\ \ \ \ 
X(S,\Pi):=F(Y(S,\Pi)),$$
where $\rho(\Pi)$ is the element $\rho$ defined for given $\Pi$.
Note that $X=X(S,\Pi)$ for the corresponding pair $(S,\Pi)$.

\subsection{Odd reflections}\label{odd}
Recall a notion of odd reflections, see~\cite{S}. Let $\Pi$ be a set of simple
roots and $\Delta_+(\Pi)$ be the corresponding set of positive roots. 
Fix a simple isotropic root $\beta\in\Pi$ and set
$$s_{\beta}(\Delta_+):=\Delta_+(\Pi)\setminus\{\beta\}\cup\{-\beta\}.$$
For each $P\subset \Pi$ set 
$s_{\beta}(P):=\{s_{\beta}(\alpha)|\ \alpha\in P\}$, where
$$ \text{ for }\alpha\in\Pi\ \ \ \ 
s_{\beta}(\alpha):=\left\{ \begin{array}{ll}
-\alpha & \text{ if } \alpha=\beta,\\
\alpha & \text{ if } (\alpha,\beta)=0,\alpha\not=\beta\\
\alpha+\beta  & \text{ if } (\alpha,\beta)\not=0.
\end{array}\right.
$$

By~\cite{S}, $s_{\beta}(\Delta_+)$ is a set of
positive roots  (i.e., $s_{\beta}(\Delta_+)=\Delta_+(f)$ 
for some functional $f$) and the corresponding set of simple roots is
$s_{\beta}(\Pi)$. 
Clearly, $\rho(s_{\beta}(\Pi))=\rho(\Pi)+\beta$.

Let $(S,\Pi)$ be an admissible pair.
Take $\beta\in S$. Then for any $\beta'\in S\setminus\{\beta\}$ one 
has $s_{\beta}(\beta')=\beta'$ so
$s_{\beta}(S)=(S\setminus\{\beta\})\cup\{-\beta\}$.
Clearly, the pair $(s_{\beta}(S),s_{\beta}(\Pi))$ 
is  admissible. Since $\rho(s_{\beta}(\Pi))=\rho(\Pi)+\beta$, one has
$Y(S,\Pi)=Y(s_{\beta}(S),s_{\beta}(\Pi))$.

\subsection{}
\begin{lem}{lemnewS}
Assume that $\gamma,\gamma'$ are isotropic roots such that
$$\gamma\in S,\ \gamma'\in \Pi,\ 
\gamma+\gamma'\in\Delta^{\#},\  \ (\gamma', \beta)=0
\text{ for each }\beta\in S\setminus\{\gamma\}.$$
Then the pair $(S',\Pi)$, where
$S':=(S\cup\{\gamma'\})\setminus\{\gamma\}$, is
admissible  and $X(S,\Pi)=X(S',\Pi)$.
\end{lem}
\begin{proof}
It is clear that the pair $(S',\Pi)$ is admissible. Set
$\alpha:=\gamma+\gamma'$ and let $s_{\alpha}\in W^{\#}$ be 
the reflection with respect to the root $\alpha$. 
Our assumptions  imply that
\begin{equation}\label{salli}
s_{\alpha}\rho=\rho;\ \  \ s_{\alpha}\gamma'=-\gamma;\ \ \ 
s_{\alpha}\beta=\beta \text{ for }
\beta\in S\setminus\{\gamma\}.
\end{equation}
Therefore 
$s_{\alpha}(Y(S',\Pi))=Y(S,\Pi)e^{-\gamma}$ that is 
$F(Y(S',\Pi))=F\bigl(s_{\alpha}(Y(S,\Pi)e^{-\gamma})\bigr)$.
Since $F\circ s_{\alpha}=-F$, the required formula $X(S,\Pi)=X(S',\Pi)$
is equivalent to the equality 
$F\bigl(Y(S,\Pi)(1+e^{-\gamma})\bigr)=0$, which follows from the fact
that the expression
$$Y(S,\Pi)(1+e^{-\gamma})=\frac{e^{\rho}}
{\prod_{\beta\in S\setminus\{\gamma\}}(1+e^{-\beta})}$$
is, by~(\ref{salli}), $s_{\alpha}$-invariant.
\end{proof}

\subsection{Equivalence classes}\label{fPi}
We consider the types $A(m-1,n-1), B(m,n), D(m,n)$ with all possible $m,n$.
We express the roots in terms of linear functions $\xi_1,\ldots,\xi_{m+n}$,
see~\cite{Ksuper}, such that
$$\begin{tabular}{|l || l | l| l|}
\hline
$\Delta$ & A(m-1,n-1) & B(m,n) & D(m,n)\\
\hline
$\Delta_{+,0}$ & $U$ & $U'\cup\{\xi_i\}_{i=1}^m\cup
\{2\xi_i\}_{i=m+1}^{m+n}$ & $U'\cup
\{2\xi_i\}_{i=m+1}^{m+n}$\\
\hline
$\Delta_1$ & $\{\pm(\xi_i-\xi_j)\}_{1\leq i\leq m<j\leq m+n}$ &
$\{\pm\xi_i\pm\xi_j;\pm \xi_j)\}_{1\leq i\leq m<j\leq m+n}$ &
$\{\pm\xi_i\pm\xi_j\}_{1\leq i\leq m<j\leq m+n}$ \\
\hline
\end{tabular}$$
and
$$\begin{array}{l}
U:=\{\xi_i-\xi_j|\ 1<i<j\leq m\text{ or }m+1\leq i<j\leq m+n\},\\
U':=\{\xi_i\pm\xi_j|\ 1<i<j\leq m\text{ or }m+1\leq i<j\leq m+n\}.\end{array}$$

\subsubsection{}
For a system of simple roots  $\Pi\subset \spn\{\xi_i\}_{i=1}^{m+n}$ take
an element $f_{\Pi}\in \spn\{\xi_i^*\}_{i=1}^{m+n}$
such that $\langle f_{\Pi},\alpha\rangle=1$ 
for each $\alpha\in\Pi$ (the existence of $f_{\Pi}$ follows from linear 
independence of the elements in $\Pi$). Note that $f_{\Pi}$ is unique
if $\fg$ is not of the type $A(m-1,n-1)$; for $A(m-1,n-1)$ we fix $f_{\Pi}$ by
 the additional condition 
$\min_{1\leq i\leq m+n}\langle f_{\Pi},\xi_i\rangle=1$. In
the notation of Sect.~\ref{intro} one has $\Pi=\Pi(f_{\Pi})$. 
We will use the following properties of $f_{\Pi}$:

(i)  $\langle f_{\Pi},\alpha\rangle\in\mathbb{Z}\setminus\{0\}$
for all $\alpha\in\Delta$; 

(ii) if $\alpha\in\Delta^+$, then $\langle f_{\Pi},\alpha\rangle\geq 1$;

(iii) a root $\alpha\in\Delta$ is simple iff 
$\langle f_{\Pi},\alpha\rangle=1$.

Write $f_{\Pi}=\sum_{i=1}^{m+n} x_i\xi_i$.
By (i),  the $x_i$s are pairwise  different; by (ii) ,
$x_i\pm x_{i+1}>0$ for $i\not=m$ ($1\leq i\leq m+n-1$);
by (iii), a root $\xi_i-\xi_j$ is simple
iff $x_i-x_j=1$. 

\subsubsection{}\label{xis}
For type $A(m-1,n-1)$ all roots are of the form 
$\xi_i-\xi_j$. In particular, $\{x_i\}_{i=1}^{m+n}=\{1,\ldots, m+n\}$.

Consider the case $B(m,n)$. In this case for each $i$ one has
$\xi_i\in\Delta_{+}$  so $x_i\geq 1$. Therefore, by (iii), 
a simple root can not be of the form $\pm(\xi_i+\xi_j)$ so 
$\Pi=\{\xi_{i_1}-\xi_{i_2},\xi_{i_2}-\xi_{i_3},\ldots,
\xi_{i_{m+n-1}}-\xi_{i_{m+n}},\xi_{i_{m+n}}\}$ and
$\{x_i\}_{i=1}^{m+n}=\{1,\ldots, m+n\}$.

\subsubsection{}
Consider the case $D(m,n)$. Using (i), (ii) and the fact that 
$2\xi_{m+n}\in\Delta_{+,0}$, we conclude
$$\begin{array}{l}
\forall j\ 2x_j\in\mathbb{Z};\ \ \ \ \ 
\forall i\not=j\ x_i\pm x_j\in\mathbb{Z}\setminus\{0\};\\
x_i-x_j\geq j-i\text{ for }i<j\leq m\text{ or } m<i<j;\\
  x_j>0 \text{ for }j>m;\ \ \ x_j+x_i>0\text{ for }i<j\leq m.
\end{array}$$

Recall that, if $-(\xi_p+\xi_q)\in \Pi\cap\Delta_1$,
then  $\Pi':=s_{-\xi_p-\xi_q}(\Pi)$ contains
$\xi_p+\xi_q$. Let us {\em assume
 that } $\xi_p+\xi_q\in\Pi\cap\Delta_1, p<q$, that is 
$x_p+x_q=1,\ p\leq m<q$. If $p<m$, then 
$x_p+x_q>\pm (x_m+x_q)$
(because $x_p\pm x_m, 2x_q>0$), so $x_p+x_q>|x_m+x_q|\geq 1$, a contradiction.
Hence $p=m$, in particular,
\begin{equation}\label{xm}
\pm (\xi_p+\xi_q)\in\Pi\cap\Delta_1,\ p<q\ \Longrightarrow\ p=m.
\end{equation}
Since $2x_q\in\mathbb{Z}_{>0},\ x_p+x_q=1$ and $x_q\not=x_m$, one has
$x_m\leq 0$. Therefore the assumption implies
$$x_m+x_q=1 \ \& \ q>m \ \&\  x_m\leq 0.$$

Since $\xi_{m-1}-\xi_m\in\Delta^+$ 
there exists a simple root of the form $\pm\xi_s-\xi_m$ ($s\not=m$).

First, consider the case when $\xi_s-\xi_m\in\Pi$, that is $x_s-x_m=1$.
Since $x_m\leq 0$, one has $x_s+x_m\leq x_s-x_m=1$ so
$x_s+x_m<0$, that is $-(\xi_s+\xi_m)\in \Delta_+$.
Therefore $s>m$ and $x_q+x_s=2$, because
$x_q+x_m=x_s-x_m=1$. Since $x_{m+n-i}\geq 1/2+i$ for
$i<n$, we conclude that either $x_q=3/2,x_m=-1/2,x_s=1/2$, which contradicts
to $x_m+x_s\not=0$, or $q=s=m+n, x_{m+n}=1, x_m=0$. Hence $\xi_s-\xi_m\in\Pi$
implies $q=m+n, x_{m+n}=1, x_m=0$ and $\xi_{m+n}\pm\xi_m\in\Pi$.

Now consider the case when $-\xi_s-\xi_m\in\Pi$. Then $s>m$ and
$x_m+x_q=-x_m-x_s=1$ so $x_q-x_s=2$. Since $x_q-x_{q+i}\geq i$
one has $s=q+1$ or $s=q+2$. If $s=q+2$, then  we have $2=x_q-x_{q+2}
=(x_q-x_{q+1})+(x_{q+1}-x_{q+2})$ that is $x_q-x_{q+1}=1$ so
$x_m+x_{q+1}=0$, a contradiction. Hence $s=q+1$ that is $-x_m-x_{q+1}=1$.
For $i<m$ one has $0<x_i+x_m=-x_{q+1}-1+x_i$ so $1<x_i-x_{q+1}$.
Therefore for $i<m,t\geq q+1$ one has $x_i+x_t>x_i-x_t>1$, so  the roots
$\pm\xi_i\pm\xi_t$ are not simple.  Assume that $m\geq n$ and 
$(S,\Pi)$ is an admissible pair.
Then for each $m<t\leq m+n$ there exists $i_t<m$ 
such that one of the roots
$\pm\xi_t\pm\xi_{i_t}$ is simple (and lies in $S$)
and $i_t\not=i_p$ for $t\not=p$.
By above, $i_{q+1}=m$ and there is no suitable $i_t$ for $t>q+1$.
Hence $q+1=m+n$ and $(-\xi_m-\xi_{m+n})\in S$, that
is $\xi_m+\xi_{m+n}\in s_{-\xi_m-\xi_{m+n}} S$.

\subsubsection{}
We conclude that if $(S,\Pi)$ is an admissible pair for $\fg=D(m,n)$, 
then one of the following possibilities hold: either 
all elements of $S$ are of the form $\xi_i-\xi_j$, or 
all elements of $S$ except one are of the form $\xi_i-\xi_j$, 
and this exceptional one is $\beta$, where

(1) $\beta=\xi_m+\xi_{m+n}$ and $\xi_{m+n}-\xi_m\in\Pi$;

(2) $m<n$ and  $\beta=\xi_m+\xi_{s}, m<s<m+n$, 
$-\xi_{s+1}-\xi_m\in\Pi$;

(3) $\beta:=-(\xi_m+\xi_s)$, the pair $(s_{\beta}S,s_{\beta}\Pi)$
is one of whose described in (1)-(2).

\subsubsection{}
Consider the case $D(m,n),\ n\geq m$. Then 
$\Delta^{\#}=\{\xi_i\pm\xi_j;2\xi_i\}_{i=m+1}^{m+n}$. Let $(S,\Pi)$ be an
admissible pair.
If $\xi_m+\xi_s\in S$ for $s<m+n$, then, by above,
$-(\xi_m+\xi_{s+1})\in\Pi$  and
the pair $(S,\Pi)$ is equivalent to the pair 
$((S\setminus\{\xi_m+\xi_s\})\cup\{-\xi_{m}-\xi_{s+1}\},\Pi)$, which
is equivalent to the pair $((S\setminus\{\xi_m+\xi_s\})
\cup\{\xi_{m}+\xi_{s+1}\},s_{-\xi_{m}-\xi_{s+1}}\Pi)$.
Thus a pair $(S,\Pi)$ with $\xi_m+\xi_s\in S$
is equivalent to a pair  $(S',\Pi')$ with $\xi_m+\xi_{m+n}\in S$.
If $\xi_m+\xi_{m+n}\in S$, then, by above, $\xi_{m+n}-\xi_m\in \Pi$ 
and the pair $(S,\Pi)$ is equivalent to the pair 
$((S\setminus\{\xi_{m+n}+\xi_m\})\cup\{\xi_{m+n}-\xi_m\},\Pi)$. 
We conclude that any pair $(S,\Pi)$ is equivalent to a pair 
$(S',\Pi')$, where $S'=\{\xi_i-\xi_{i_j}\}_{i=1}^m$.

Consider the case $D(m,n),\ m>n$. Then 
$\Delta^{\#}=\{\xi_i\pm\xi_j\}_{i=1}^{m}$. By above,
any pair $(S,\Pi)$ is equivalent either to a pair 
$(S',\Pi')$, where $S'=\{\xi_i-\xi_{i_j}\}_{i=1}^n$, or
to a pair $(S',\Pi')$, where $S'=\{\xi_i-\xi_{i_j}\}_{i=1}^{n-1}\cup
\{\xi_{m+n}+\xi_m\}$ and $\xi_{m+n}-\xi_m\in \Pi'$.

\subsubsection{}
Let $(S,\Pi)$ be an admissible pair.
We conclude that any pair $(S,\Pi)$ is equivalent to a pair $(S',\Pi')$, where
either $S'=\{\xi_i-\xi_{j_i}\}_{i=1}^{\min(m,n)}$, or,
for $D(m,n),\ m>n$ $S'=\{\xi_i-\xi_{i_j}\}_{i=1}^{n-1}\cup
\{\xi_{m+n}+\xi_m\}$ and $\xi_{m+n}-\xi_m\in \Pi'$.

\subsection{}\label{fPi2}
Fix a set of simple roots of $\fg$ and construct $f_{\Pi}$ as
in~\ref{fPi}. We mark the points $x_i$ on the real line by $a$'s and $b$'s 
in one of the following ways: 

(M) mark $x_i$ by $a$ (resp., by $b$) if $1\leq i\leq m$ 
(resp., if $m<i\leq m+n$);

(N) mark $x_i$ by $b$ (resp., by $a$) if $1\leq i\leq m$ 
(resp., if $m<i\leq m+n$).

We use the marking (M) if $\Delta^{\#}$ lies in the span
of $\{\xi_i\}_{i=1}^m$ and the marking (N) if $\Delta^{\#}$ lies in the span
of $\{\xi_i\}_{i=m+1}^{m+n}$. Note that in all cases
the number of $a$s is not smaller than the number of $b$s.

We fix $\Delta^{\#}$ and an admissible pair $(S,\Pi)$ such that
$S=\{\xi_i-\xi_{j_i}\}_{i=1}^{\min(m,n)}$, or
for $D(m,n), m>n$, $S=\{\xi_i-\xi_{i_j}\}_{i=1}^{n-1}\cup
\{\xi_{m+n}+\xi_m\}$.

If $\xi_i-\xi_j\in S$ (resp., $\xi_i+\xi_j\in S$) we 
draw a bow $\smile$ (resp., $\frown$) between the points $x_i$ and $x_j$.
 Observe that the points connected by a bow are neighbours and 
they are marked by different letters ($a$ and $b$). We say that a marked point
is a vertex if it is a vertex of a bow. Note that the
bows do not have common vertices and that 
all  points marked by $b$ are vertices. 

From now one we consider the diagrams, which are
sequence of $a$s and $b$s endowed with the bows
(we do not care about the values of $x_i$).
For example, for $\fg=A(4,1)$ and $\Pi=\{\xi_1-\xi_2;\xi_2-\xi_6;
\xi_6-\xi_7;\xi_7-\xi_3;\xi_3-\xi_4;\xi_4-\xi_5\}$ we  choose
$f=\xi_5^*+2\xi_4^*+3\xi_3^*+4\xi_7^*+5\xi_6^*+6\xi_2^*+7\xi_1^*$ and
taking $S=\{\xi_2-\xi_6;\xi_7-\xi_3\}$ we obtain the diagram 
$aaa\smile bb\smile a a$; for $\fg=B(2,2)$ and
$\Pi=\{\xi_1-\xi_3;\xi_3-\xi_2;\xi_2-\xi_4;\xi_4\}$ 
 we  choose $f=\xi_4^*+2\xi_2^*+3\xi_3^*+4\xi_1^*$
and taking $S=\{\xi_1-\xi_3;\xi_2-\xi_4\}$ we obtain the diagram 
$a\smile ba\smile b$ for the marking (N), and
$b\smile ab\smile a$ for the marking (M).
Observe that a diagram containing
$\frown$  appear only in the case $D(m,n),\ m>n$,
and such  a diagram starts from $a\frown b$, because
$\xi_{m+n}+\xi_m\in S$ forces $\xi_{m+n}-\xi_m\in\Pi$, which implies
$x_m=0, x_{m+n}=1, x_i>0$ for all $i\not=m$.

\subsubsection{}
Let us see how the odd reflections and
the operations of the second type, introduced in~\Lem{lemnewS},
change our diagrams.

Recall that $\xi_i-\xi_j\in\Delta_+$ iff $x_i>x_j$.
For an odd simple root $\beta$ one has 
$s_{\beta}(\Delta_+)=(\Delta_+\setminus\{\beta\})\cup\{-\beta\}$,
so the order of $x_i$s for $s_{\xi_p-\xi_q}(\Delta_+)$ 
is obtained from  the order of $x_i$s for $\Delta_+$ by
the interchange of $x_p$ and $x_q$ (if $\xi_p-\xi_q\in \Pi\cap\Delta_1$).
Therefore the odd reflection with respect to 
$\xi_p-\xi_q\in S$ corresponds to the following operation with
the diagram: we interchange  the
vertices (i.e., the marks $a,b$) of the corresponding bow:

$$...a\smile b...\ \mapsto ...b\smile a...;\ \ \ \ \ 
...b\smile a...\ \mapsto ...a\smile b...$$

If the diagram has a part $a\smile ba$ (resp. $ab\smile a$), 
where the last (resp. the first) sign $a$ is not a vertex,
 and $x_i,x_j,x_k$ are the
corresponding points, then the quadrapole
$(S,\Pi), \gamma:=\xi_i-\xi_j,\gamma':=\xi_j-\xi_k$ 
(resp. $\gamma':=\xi_i-\xi_j,\gamma:=\xi_j-\xi_k$)
satisfies the assumptions of~\Lem{lemnewS}.
The operation $(S,\Pi)\mapsto (S',\Pi)$, where $S':=(S\setminus\{\gamma\})\cup
\{\gamma'\}$ corresponds to the  following operation with our diagram: 
$a\smile ba\mapsto ab\smile a$ (resp. 
$ab\smile a\mapsto a\smile ba$). Hence we can perform
the operation of the second type
if $a,b,a$ are neighbouring points, $b$ is connected by $\smile$
with one of $a$s and another $a$ is not a vertex; in this case,
we remove the bow and connect $b$ with another $a$:
$$...ab\smile a...\ \mapsto ...a\smile ba...;\ \ \ \ \ 
...a\smile b a...\ \mapsto ...a b\smile a....$$

Since both our operations are involutions, we can consider the orbit of
a given diagram with respect to the action of the group generated 
by these operations.
Let us show that all diagrams without $\frown$ lie in the same orbit.
Indeed, using the operations $a\smile b\mapsto b\smile a$ and
$ab\smile a\mapsto a\smile ba$ we put $b$ to the first place
so our new diagram starts from $b\smile a$. 
Then we do the same with the rest of 
the diagram and so on. Finally, we obtain the diagram of the form
$b\smile ab\smile a\ldots b\smile a\ldots a$. By the same argument,
all the diagram starting from $b\frown a$ lie in the same orbit.
Hence, for $\fg\not=D(m,n), m>n$ all diagrams lie in the same orbit, and
  for $\fg=D(m,n), m>n$ there are two orbits:
the diagrams with $\frown$ and the diagrams without $\frown$.

\subsubsection{}\label{PiPi'}
We claim that if $(S,\Pi), (S,\Pi')$ are admissible pairs, then
$\Pi=\Pi'$. It is clear that if the claim is valid for some $S$, then it 
is valid for all $S'$
such that $(S,\Pi)$ is equivalent to $(S',\Pi')$. Thus
it is enough to verify the claim for any representative of the orbit.
Each $S$ determines the diagram and the diagram
determines the order of $x_i$'s (for instance,
the diagram $b\smile ab\smile a\ldots b\smile aa\ldots a$ gives
$x_{m+n}<x_m<x_{m+n-1}<x_{m-1}<\ldots<x_{m+1}<x_{m-n+1}<x_{m-n}<\ldots<x_1$
for the marking (M) and
$x_m<x_{m+n}<x_{m-1}<x_{m+n-1}<\ldots<x_1<x_{n+1}<x_{n}<\ldots<x_{m+1}$
for the marking (N)). For $A(m-1,n-1), B(m,n)$ one has 
$\{x_i\}_{i=1}^{m+n}=\{1,\ldots,m+n\}$, by~\ref{xis},
so each diagram determines $\Pi$. 

For $D(m,n),\ m>n$ the above diagram 
means that $-\xi_{m+n}+\xi_m,-\xi_m+\xi_{m+n-1},\ldots\in \Pi$.
Since $2\xi_{m+n}\in\Delta_+$, we conclude that $2\xi_{m+n}\in\Pi$ that
is $x_{m+n}=\frac{1}{2}$ and 
$\{x_i\}_{i=1}^{m+n}=\{\frac{1}{2},\ldots,m+n-\frac{1}{2}\}$. Hence 
the above diagram determines $\Pi$. 

For $D(m,n), m\leq n$ the same reasoning shows that the diagram 
$a\smile bb\smile a\ldots b\smile aa\ldots a$ determines $\Pi$.

It remains to consider the case $D(m,n),\ m>n$ and the diagram 
$a\frown b b\smile ab\smile a\ldots b\smile aa\ldots a$.
In this case, $\xi_m+\xi_{m+n}\in S$ so, by above, $x_m=0$
and  $\{x_i\}_{i=1}^{m+n}=\{0,\ldots,m+n-1\}$. Thus  the diagram
determines $\Pi$.

\subsubsection{}
Consider the case $\fg\not=D(m,n), m>n$. Fix an admissible pair $(S,\Pi)$.
By~\ref{fPi2}, any admissible pair $(S',\Pi')$ is equivalent to an 
admissible pair $(S,\Pi'')$; by~\ref{PiPi'} one has $\Pi=\Pi''$ so
$(S',\Pi')$ is equivalent to $(S,\Pi)$.

Consider the case $\fg=D(m,n), m>n$. Fix admissible  pairs $(S,\Pi),
\ (S',\Pi')$, where $S$ consists of the roots of the form $\xi_i-\xi_j$ and
$S'$  contains $\xi_m+\xi_{m+n}$. Arguing as above, we conclude that
any admissible pair $(S'',\Pi'')$ is equivalent either to $(S,\Pi)$ or
to $(S',\Pi')$.

\subsection{}
\begin{cor}{cornews}
(i) If $(S,\Pi),\ (S,\Pi')$ are admissible pairs, then
$\Pi=\Pi'$.

(ii) For $\fg\not=D(m,n), m>n$ there is one equivalence class of the pairs
$(S,\Pi)$ and the left-hand (resp., right-hand) side of~(\ref{denomKW}) 
is the same for all admissible pairs $(S,\Pi)$.

(iii) For $\fg=D(m,n), m>n$ there are two equivalence classes of the pairs
$(S,\Pi)$. In the first class $S$ consists of the elements of the form
$\pm(\xi_i-\xi_j)$ and in the second class $S$ contains a unique
element of the form $\pm(\xi_i+\xi_j)$. The left-hand side 
(resp., right-hand) of~(\ref{denomKW}) is the same for all admissible pairs
$(S,\Pi)$ belonging to the same class.
\end{cor}

\section{Steps (ii), (ii')}

\subsection{}
Assume that 
\begin{equation}\label{alpha>0}\begin{array}{l}
\forall\alpha\in\Pi\ \ (\alpha,\alpha)\geq 0;\\ 
W^{\#}\text{ is generated by the set of reflections } 
\{s_{\alpha}|\ (\alpha,\alpha)>0\}.
\end{array}
\end{equation}

We start from the following lemma.
\subsubsection{}
\begin{lem}{lem1}
One has

(i) $\rho\geq w\rho$ for all $w\in W^{\#}$;

(ii) the stabilizer of $\rho$ in $W^{\#}$  is generated
by the set $\{s_{\alpha}|\ (\alpha,\alpha)>0,\ (\alpha,\rho)=0\}$.
\end{lem}
\begin{proof}
Since $(\alpha,\rho)=\frac{1}{2}(\alpha,\alpha)\geq 0$ for all 
$\alpha\in\Pi$, one has $(\beta,\rho)\geq 0$ for all $\beta\in\Delta_+$.
Take $w\in W^{\#}, w\not=\id$. One has
$w=w's_{\beta}$, where $w'\in W^{\#},\beta\in\Delta_+$ are such that
the length of $w$ is greater than the length of $w'$ 
and $w'\beta\in\Delta_+$ (see, for example, \cite{Jbook}, A.1). One has
$$\rho-w\rho=\rho-w'\rho+\frac{2(\rho,\beta)}
{(\beta,\beta)}\cdot(w'\beta).$$
Now (i) follows by induction on the length of $w$, since
$(\rho,\beta)\geq 0,(\beta,\beta)>0,\ w'\beta\in\Delta_+$.
For (ii), note that  $\rho\geq w'\rho$ by (i), thus $w\rho=\rho$ forces 
$\rho=w'\rho$ and $(\rho,\beta)=0$. Hence (ii) also follows by induction
on the length of $w$.
\end{proof}

\subsubsection{}\label{supprho}
Retain notation of~(\ref{expX}). For $w\in W^{\#}$ one has
$-\varphi(w),|w|\mu\in Q^+$, by definition, and $w\rho\leq \rho$,
by~\Lem{lem1}. Therefore $\varphi(w)-|w|\mu+w\rho\leq \rho$ 
and the equality means that $\mu=0,\ w\rho=\rho$ and  
$\varphi(w)=0$, that is $wS\subset\Delta_+$.
The above inequality gives $\supp(X)\subset (\rho-Q^+)$. 
Moreover, by above, the coefficient of $e^{\rho}$ in the expansion
of $X$ is equal to 
$$\sum_{w\in W^{\#}: w\rho=\rho,\ wS\subset\Delta_+} \sgn(w).$$

\subsection{Root systems}\label{Rootsy}
Recall that we consider all choices of $\Delta_+$ with 
a fixed $\Delta_{+,0}$. 
From now on we {\em assume that $m\geq n$} (we consider the types
$A(m,n)$, $B(m,n)$, $B(n,m)$, $D(m,n)$, $D(n,m)$) and
we embed our root systems in the standard lattices spanned by
$\{\vareps_i,\delta_j: 1\leq i\leq m, 1\leq j\leq n\}$ chosen
in such a way that
$\Delta^{\#}=\Delta_0\cap\spn\{\vareps_i\}_{i=1}^m$. 
More precisely, for $A(m,n), m\geq n$ we take
$$
\Delta_{+,0}=\{\vareps_i-\vareps_{i'};\delta_j-\delta_{j'}| 1\leq i<i'\leq m,
1\leq j<j'\leq n\},\ \ \ \Delta_1=\{\pm(\vareps_i-\delta_j)\};
$$
for other cases we put
$$U':=\{\vareps_i\pm\vareps_{i'};\delta_j\pm\delta_{j'}| 1\leq i<i'\leq m,
1\leq j<j'\leq n\}$$
and then 
$$\begin{array}{lll}
\Delta_{+,0}=U'\cup \{\vareps_i\}_{i=1}^m\cup\{2\delta_j\}_{j=1}^n, &
\Delta_1=\{\pm\vareps_i\pm\delta_j,\pm\delta_j\} & 
\text{ for }B(m,n), m>n \\
& & \text{ and }B(n,n), \Delta^{\#}=B(n);\\
\Delta_{+,0}=U'\cup \{2\vareps_i\}_{i=1}^m\cup\{\delta_j\}_{j=1}^n, &
\Delta_1=\{\pm\vareps_i\pm\delta_j,\pm\vareps_i\} &
\text{ for }B(n,m), m>n \\
& & \text{ and }B(n,n), \Delta^{\#}=C(n);\\
\Delta_{+,0}=U'\cup \{2\delta_j\}_{j=1}^n, &
\Delta_1=\{\pm\vareps_i\pm\delta_j\} & \text{ for }D(m,n), m>n; \\
\Delta_{+,0}=U'\cup \{2\vareps_i\}_{i=1}^m, &
 \Delta_1=\{\pm\vareps_i\pm\delta_j\} & \text{ for }D(n,m), m\geq n.
\end{array}$$

We normalize the form $(-,-)$ by the condition $(\vareps_i,\vareps_j)=
-(\delta_i,\delta_j)=\delta_{ij}$. One has
$\Delta^{\#}=\Delta_0\cap\spn\{\vareps_i\}_{i=1}^m=
\{\alpha\in\Delta_0|\ (\alpha,\alpha)>0\}$.

\subsection{Choice of $(S,\Pi)$}\label{rootsys}
For the case $B(n,n)$ set 
$$S:=\{\delta_i-\vareps_i\},\ \
\Pi:=\{\delta_1-\vareps_1,\vareps_1-\delta_2,
\delta_2-\vareps_2,\ldots, \delta_n-\vareps_n,\vareps_n\},$$
(the root $\vareps_n$ may be even or odd, depending on the choice of 
$\Delta^{\#}$). For other cases set
$$S:=\{\vareps_i-\delta_i\}_{i=1}^n.$$

In order to describe $\Pi$ introduce 
$$P:=\{\vareps_1-\delta_1,
\delta_1-\vareps_2,\vareps_2-\delta_2,\delta_2-\vareps_3,
\ldots, \vareps_n-\delta_n,\delta_{n}-\vareps_{n+1},
\vareps_{n+1}-\vareps_{n+2},\ldots, 
\vareps_{m-1}-\vareps_m\}$$
and set
$$\begin{tabular}{|l || l | l| l|l| l|}
\hline
$\Delta$ & A(m-1,n-1) & B(m,n), B(n,m), $n>m$ & D(n,m), $m>n$ & D(m,n) $m>n$\\ 
\hline
$\Pi$ & $P$ & $P\cup \{\vareps_m\}$ & $P\cup \{2\vareps_m\}$ &
$P\cup\{\vareps_{m-1}+\vareps_{m}\}$\\
\hline
\end{tabular}$$
$$\Pi:=\{\vareps_1-\delta_1,
\delta_1-\vareps_2,\vareps_2-\delta_2,\delta_2-\vareps_3,
\ldots, \vareps_n-\delta_n,\vareps_n+\delta_n\}\text { for } D(n,n).$$

Note that the assumptions~(\ref{alpha>0}) hold.

\subsection{Step (ii): the case $(\vareps_m+\delta_n)\not\in\Pi$}

By~\ref{supprho}, $\supp(X)\subset (\rho-Q^+)$ and 
the coefficient of $e^{\rho}$ in the expansion
of $X$ is $\sum_{w\in W^{\#}: w\rho=\rho,\ wS\subset\Delta_+} \sgn(w)$.
Thus it is enough to show that
\begin{equation}\label{step4}
w\in W^{\#} \text{ s.t. } 
wS\subset\Delta_+,\ w\rho=\rho\ \Longrightarrow\ w=\id.
\end{equation}

\subsubsection{}\label{Sk}
Since $(\alpha,\alpha)\geq 0$ for each $\alpha\in\Pi$, one has
$(\rho,\beta)=0$ for $\beta\in\Delta$
iff $\beta$ is a linear combination of isotropic simple
roots. 

Consider the case $\fg\not=D(n,n)$ (one has $\rho=0$ for $D(n,n)$).
In this case $(\rho,\beta)=0$ for $\beta\in\Delta^{\#}_+$ forces
$\beta=\vareps_i-\vareps_j$ for $i<j\leq \min(m,n+1)$.
From~\Lem{lem1} we conclude
\begin{equation}\label{eqlem1}
\text{ for } \fg\not=D(n,n)\ \ \ \Stab_{W^{\#}}\rho=\left\{ 
\begin{array}{ll} S_n, &\text{ if } m=n,\\
S_{n+1}, &\text{ if } m>n,
\end{array}
\right.\end{equation}
where $S_k\subset W^{\#}$ is the symmetric group, consisting of the
permutations of $\vareps_1,\ldots, \vareps_k$ (that is
for $w\in S_k$ one has $w\vareps_i=\vareps_{j_i}$ and
$j_i=i$ for $i>k$).

\subsubsection{}
Take $w\in \Stab_{W^{\#}}\rho$ such that $wS\subset\Delta_+$.
Let us show that $w=\id$.

Consider the case when $\fg\not=B(n,n), D(n,n)$.  
Then $S=\{\vareps_i-\delta_i\}_{i=1}^n$. Combining~(\ref{eqlem1})
and the fact that $\vareps_j-\delta_i\in\Delta_+$ 
iff $j\leq i$, we conclude that
$w\vareps_{i}=\vareps_{j_i}$ for  $j_i\leq i$
if $i\leq n$ and $j_i=i$ for $i>n+1$. Hence $w=\id$ as required.

For $\fg=D(n,n)$ one has $S=\{\vareps_i-\delta_i\}_{i=1}^n$.
Since $-\vareps_i-\delta_j\not\in \Delta_+$ for all $i,j$,
the condition $wS\subset\Delta_+$ forces
$w\in S_n$ (see~\ref{Sk} for notation). Repeating the above argument,
 we obtain $w=\id$.

For $\fg=B(n,n)$ one has $S=\{\delta_i-\vareps_i\}_{i=1}^n$.
Combining~(\ref{eqlem1}) and the fact that
$\delta_i-\vareps_j\in\Delta_+$ iff $j\geq i$, we obtain  for all
$i=1,\ldots,n$ that $w\vareps_{i}=\vareps_{j_i}$ for some $j_i\leq i$. 
Hence $w=\id$ as required. This establishes~(\ref{step4}) and (ii) 
for the case $(\vareps_m+\delta_n)\not\in\Pi$.

\subsection{Step (ii): the case $(\vareps_m+\delta_n)\in\Pi$}
Consider the case $D(m,n),\ m>n, (\vareps_m+\delta_n)\in\Pi$.
We retain notation of~\ref{Rootsy} and choose a new pair $(S,\Pi)$:
$S =\{\vareps_i-\delta_i\}_{i=1}^{n-1}\cup\{\vareps_m+\delta_n\}$
and 
$$\Pi:=\{\vareps_1-\delta_1,\delta_1-\vareps_2,\ldots,
\delta_{n-1}-\vareps_{n}\}\cup\{\vareps_{i}-\vareps_{i+1}\}_{i=n}^{m-2}
\cup\{\vareps_{m-1}-\delta_n,\delta_n-
\vareps_m,\delta_n+\vareps_m\}.$$

The assumptions~(\ref{alpha>0}) are satisfied.
By~\ref{supprho}, the coefficient of $e^{\rho}$ in the expansion
of $X$ is equal to $\sum_{w\in A} \sgn(w)$, where
$A:=\{w\in \Stab_{W^{\#}}\rho|\ wS\subset\Delta_+\}$.
Let us show that $A=\{ \id,s_{\vareps_{m-1} -\vareps_m},
s_{\vareps_{m-1}-\vareps_m}s_{\vareps_{m-1} +\vareps_m}\}$;
this implies that the coefficient of $e^{\rho}$ in the expansion
of $X$ is equal to $1$.

Take $w\in W^{\#}$ such that $wS\subset\Delta_+$ .
 Note that $-\vareps_j-\delta_i\not\in\Delta_+$ for all $i,j$ 
and for $i<n$ one has $\vareps_j-\delta_i\in\Delta_+$ 
iff $j\leq i$. The assumption $wS\subset\Delta_+$ means that
$w\vareps_{i}-\delta_i\in\Delta_+$ for all $i<n$ and 
 $w\vareps_{m}+\delta_n\in\Delta_+$.
For $i<n$ this gives $w\vareps_{i}=\vareps_{j_i}$ 
for some $j_i\leq i$. Hence $w\vareps_{i}=\vareps_i$ for $i=1,\ldots,n-1$.
The remaining condition  $w\vareps_{m}+\delta_n\in\Delta_+$
means that $w\vareps_{m}=\vareps_{j_m}$ or  $w\vareps_{m}=-\vareps_{m}$.

For $m=n+1$ one has $\rho=0$, so $w\in A$ iff $w\in W^{\#},\ 
wS\subset\Delta_+$. Thus, by above, $w\in A$ iff
 $w\vareps_{i}=\vareps_i$ for $i<n=m-1$ and
$w\vareps_{m}\in\{\pm \vareps_{m},\vareps_{m-1}\}$.

Take $m>n+1$. The roots $\beta\in\Delta^{\#}_+$ such that
$(\rho,\beta)=0$  are of the form
$\beta=\vareps_i-\vareps_j$ for $i<j\leq n$ or
$\beta=\vareps_{m-1}\pm\vareps_m$. From~\Lem{lem1} we conclude
that  the subgroup $\Stab_{W^{\#}}\rho$
is a product of $S_n$ defined in~\ref{Sk} and the group generated by
the reflections $s_{\vareps_{m-1}\pm\vareps_m}$.
By above, $w\in A$ iff $w\vareps_{i}=\vareps_i$ for $i<m-1$
and $w\vareps_{m}\in\{\pm \vareps_{m},\vareps_{m-1}\}$.

Since $W^{\#}$ is the Weyl group of $D(m)$, i.e. the group
of signed permutations of $\{\vareps_{i}\}_{i=1}^m$, changing
the even number of signs, the set
$$\{w\in W^{\#}|\ w\vareps_{i}=\vareps_i\text{ for }i<m-1 \ \&\ \ 
w\vareps_{m}\in\{\pm \vareps_{m},\vareps_{m-1}\}\}$$
is  $\{ \id,s_{\vareps_{m-1} -\vareps_m},
s_{\vareps_{m-1}-\vareps_m}s_{\vareps_{m-1} +\vareps_m}\}$.
Hence $A=\{ \id,s_{\vareps_{m-1} -\vareps_{m}},
s_{\vareps_{m-1}-\vareps_{m}}s_{\vareps_{m-1} +\vareps_{m}}\}$
as required. This establishes (ii) for the case $(\vareps_m+\delta_n)\in\Pi$.

\subsection{Step (ii')}
Consider the case $\fgl(n|n)$. One has $\rho=0$.
Set $\xi:=\sum_{\beta\in S}\beta=\sum\vareps_i-\sum\delta_i$.
Let us verify that 
\begin{equation}\label{skis}
\mathbb{Q}\xi\cap \supp (Re^{\rho}-X)=\emptyset.
\end{equation}
Indeed, it is easy see that $\xi$ has a unique presentation as a 
positive linear combination of positive roots:
\begin{equation}\label{ski}
\xi=\sum_{\alpha\in\Delta_+}m_{\alpha}\alpha, \ m_{\alpha}\geq 0\ \
\Longrightarrow\ \ m_{\beta}=1\text{ for }\beta\in S,\ m_{\alpha}=0
\text{ for }\alpha\not\in S.
\end{equation}
This implies that for $s\not\in\mathbb{Z}_{\geq 0}$
the coefficients of $e^{-s\xi}$ in $Re^{\rho}=R$ and in $X$
are equal to zero, and that the coefficient of 
$e^{-s\xi}$ in $R$ is equal to $(-1)^{sn}$ for $s\in\mathbb{Z}_{\geq 0}$. 
It remains to show that the coefficient 
of $e^{-s\xi}$ in $X$ is $(-1)^{sn}=(-1)^{\htt (s\xi)}$. 
It is enough to verify that
$|w|\mu-\varphi(w)=s\xi$ for $w\in W^{\#}$ implies $w=\id$. 
Assume that $|w|\mu-\varphi(w)=s\xi$. By definition,
$|w|\mu,-\varphi(w)\in Q^+$. From~(\ref{ski}) we conclude that
$|w|\mu,-\varphi(w)\in\mathbb{Z}_{\geq 0}S$.
Recall that $\varphi(w)=\sum_{\beta\in S: w\beta\in\Delta_-} w\beta$.
By~(\ref{ski}), $-\varphi(w)\in\mathbb{Z}_{\geq 0}S$ implies
$(-w\beta)\in S$ for each
$\beta\in S$ such that $w\beta\in\Delta_-$. However, 
$(-w\beta)\not\in S$ for any $w\in  W^{\#},\ \beta\in S$, because
 $-w(\vareps_i-\delta_i)=\delta_i-w\vareps_i\not\in S$.
Thus  $wS\subset\Delta_+$ that is $w\vareps_i=\vareps_{i_j}$
for $i_j\leq i$. Hence $w=\id$. This establishes~(\ref{skis}).

\section{$W$-invariance: step (iii)}
\label{Winv}
In this section we prove that $X$ defined by the formula~(\ref{defX})
is $W$-skew-invariant for a certain admissible pair $(S,\Pi)$;
for the case $D(m,n)\ m>n$ we prove this for two admissible pairs $(S,\Pi)$
and $(S',\Pi')$, which are representatives of the equivalence classes
defined in Sect.~\ref{sect(i)}.

Recall that $X$ is $W^{\#}$-skew-invariant and that
$\Delta=\Delta^{\#}\coprod\Delta_2$, that is $W=W^{\#}\times W_2$.

\subsection{Operator $F$}\label{F}
Recall the operator $F:\cR\to\cR$ given by the formula
$F(Y):=\sum_{w\in W^{\#}}\sgn(w) wY$. 
Clearly, $w(F(Y))=F(wY)$ for $w\in W_2$ and
$F(wY)=w(F(Y))=\sgn(w) F(Y)$ for $w\in W^{\#}$.
In particular, $F(Y)=0$ if $wY=Y$
for some $w\in W^{\#}$ with $\sgn(w)=-1$. One has 
$$X=F(Y),\ \text{ where }
 Y:=\frac{e^{\rho}}{\prod_{\beta\in S}(1+e^{-\beta})}.$$

Suppose that $B\in\cR$ is such that $w_2w_1 B=B$ for some $w_1\in W^{\#},
w_2\in W_2$, where $\sgn (w_1w_2)=1$. Then
$$w_2^{-1}F(B)=F(w_2^{-1}B)=F(w_1B)=\sgn(w_1) F(B),$$
that is $w_2F(B)=\sgn(w_2)F(B)$.

As a result, in order to verify $W$-skew-invariance of $F(B)$ for an
arbitrary $B\in\cR$, it is enough to show that for each generator
$y$ of $W_2$ there exists $z\in W^{\#}$ such that $\sgn (yz)=1$
and $yzB=B$ (we consider $y$ running through a set of generators of $W_2$).

\subsection{Root systems}\label{rootsystem}
We retain notation of~\ref{Rootsy} for $\Delta_{0,+}$ and $\Delta_1$,
but, except for $A(m-1,n-1)$ we do not choose the same pairs $(S,\Pi)$
as in~\ref{rootsys}. 

For $A(m-1,n-1)$ we choose $S,\Pi$ as in~\ref{rootsys} 
($S:=\{\vareps_i-\delta_i\}$).
For other cases we choose 
$S:=\{\delta_{n-i}-\vareps_{m-i}\}_{i=0}^{n-1}$ and
$$\begin{array}{ll}
\Pi:=P\cup\{\vareps_{m}\} &\text{ for }B(m,n), B(n,m) \\
\Pi:=P\cup\{\vareps_{m}+\delta_n\}&\text{ for }D(m,n)\ m>n,\\
\Pi:=P\cup\{2\vareps_{m}\},&\text{ for }D(n,m)\ m\geq n,
\end{array}$$
where
$P:=\{\vareps_1-\vareps_2,\ldots,\vareps_{m-n-1}-\vareps_{m-n},
\vareps_{m-n}-\delta_1,\delta_1-\vareps_{m-n+1},
\vareps_{m-n+1}-\delta_2,\ldots, \delta_n-\vareps_{m}\}$ for $m>n$, 
and $P:=\{\delta_1-\vareps_{1},\vareps_{1}-\delta_2,\ldots, 
\delta_n-\vareps_{n}\}$ for $m=n$. For $D(m,n), m>n$ case we consider
two admissible pairs: $(S,\Pi)$ and $(S',\Pi)$, where 
$S':=\{\delta_{n-i}-\vareps_{m-i}\}_{i=1}^{n-1}\cup\{\delta_n+\vareps_m\}$.

Recall that 
$(\vareps_i,\vareps_j)=-(\delta_i,\delta_j)=\delta_{ij}$ and notice
that $(\alpha,\alpha)\geq 0$ for all $\alpha\in\Pi$.

\subsection{$S_n$-invariance}\label{Sninv}
Let $S_n\subset W_2$ be the group of permutations of 
$\delta_1,\ldots,\delta_n$.
In all cases $S$ is of the form
$S=\pm\{\delta_i-\vareps_{r+i}\}$ for $r=0$ or $r=m-n$.
For $i=1,\ldots, n-1$ one has $(\rho,\delta_i-\delta_{i+1})=
(\rho,\vareps_{r+i}-\vareps_{r+i+1})=0$. Therefore the reflections
$s_{\vareps_{r+i}-\vareps_{r+i+1}}, s_{\delta_i-\delta_{i+1}}$
stabilize $\rho$. Since $s_{\vareps_{r+i}-\vareps_{r+i+1}}
s_{\delta_i-\delta_{i+1}}$ stabilizes the elements of $S$, one has
$s_{\vareps_{r+i}-\vareps_{r+i+1}}s_{\delta_i-\delta_{i+1}}Y=Y$
for $i=1,\ldots,n-1$. Using~\ref{F} we conclude that 
$X$ is $S_n$-skew-invariant.
In particular, this establishes $W$-invariance of $X$ for $A(m,n)$-case.

For $D(m,n), m>n$ case consider the admissible pair $(S',\Pi)$. 
Arguing as above,
one sees that $s_{\delta_i-\delta_{i+1}}(X)=-X$
for $i=1,\ldots,n-2$. Since $\delta_{n-1}-\vareps_{m-1},
\delta_n+\vareps_m\in S'$ the product
$w:=s_{\vareps_{m-1}+\vareps_{m}}s_{\delta_{n-1}-\delta_{n}}$ 
stabilizes the elements of $S'$. Since $(\rho,\delta_i)=
(\rho,\vareps_{m-n+i})=0$ for $i=1,\ldots,n$,
$w$ stabilizes $\rho$. Thus $wY=Y$ and so $s_{\delta_{n-1}-\delta_{n}}X=-X$,
by~\ref{F}. Hence $X$ is $S_n$-skew-invariant.

\subsection{$B(m,n), B(n,m), D(m,n), m>n$ cases}
In this case $W_2$ is the group of signed permutations
of $\{\delta_i\}_{i=1}^n$ so it is
generated by $s_{\delta_n}$ and the elements of $S_n$.
In the light of~\ref{F} and~\ref{Sninv},
 it is enough to verify that $s_{\delta_n}s_{\vareps_m}Y=Y$.
Set
$$\beta:=\delta_n-\vareps_m\in S.$$

Consider the cases $B(m,n), B(n,m)$.
In this case $W^{\#}$  is the group of signed permutations
of $\{\vareps_i\}_{i=1}^m$. Since $\beta,\vareps_m\in\Pi$, one has 
$(\rho, \vareps_m)=(\rho, \delta_n)=\frac{1}{2}$ so
$s_{\delta_n}s_{\vareps_m}\rho=\rho+\beta$. Clearly, 
$s_{\delta_n}s_{\vareps_m}$ 
stabilizes the elements of $S\setminus\{\beta\}$, 
and 
$s_{\delta_n}s_{\vareps_m}\beta=-\beta$. As a result,
$s_{\delta_n}s_{\vareps_m}Y=Y$ as required.

Consider the case $D(m,n), m>n$.
In this case $W^{\#}$ is the group of signed permutations of 
$\{\vareps_i\}$, which change even number of signs.
Notice that $s_{\vareps_{m-n}}s_{\vareps_{m}}\in W^{\#}$ and
$\sgn(s_{\vareps_{m-n}}s_{\vareps_{m}})=1$.
Set $w:=s_{\vareps_{m-n}}s_{\vareps_{m}}s_{\delta_n}$.
One has $(\rho,\delta_n)=(\rho,\vareps_m)=(\rho,\vareps_{m-n})=0$
so $w\rho=\rho$. Since $w$ stabilizes the elements 
of $S\setminus\{\beta\}$, one has $wY=e^{-\beta}Y$.
We obtain
$$s_{\delta_n}F(Y)=s_{\delta_n}F(s_{\vareps_{m-n}}s_{\vareps_{m}}Y)=
F(s_{\delta_n}s_{\vareps_{m-n}}s_{\vareps_{m}}Y)=F(wY)=
F(e^{-\beta}Y)$$ 
and so
$$(1+s_{\delta_n})F(Y)=F\bigl((1+e^{-\beta})Y\bigr)
=F\bigl(\frac{e^{\rho}}{\prod_{\beta'\in 
S\setminus\{\beta\}}(1+e^{-\beta'})}\bigr)=0,$$
where the last equality follows from the fact that
the reflection $s_{\vareps_{m-n}-\vareps_{m}}$ stabilizes $\rho$ and 
$S\setminus\{\beta\}$. Hence $(1+s_{\delta_n})F(Y)=0$ as required.

For the admissible pair $(S',\Pi)$ we obtain the required formula
 $(1+s_{\delta_n})F(Y)=0$ along the same lines substituting $\beta$ by
$\delta_n+\vareps_{m}$.

\subsection{Case $D(n,m), m\geq n$}
In this case $W^{\#}$ is the group of signed permutations of 
$\{\vareps_i\}$,  and
$W_2$ is the group of signed permutations
of  $\{\delta_i\}_{i=1}^n$, which change even number of signs.
Note that the reflection $s_{\delta_i}$ does not lie in
$W_2$, but $s_{\delta_i}\Delta=\Delta$, so $s_{\delta_i}$
acts on $\cR$ and this action commutes with the operator $F$.
Since $W_2$ is generated by  $s_{\delta_1}s_{\delta_2}$
and the elements of $S_n$,
it is enough to verify that $s_{\delta_1}s_{\delta_2}F(Y)=F(Y)$.
Set $\beta_i:=\delta_i-\vareps_{m-n+i}\in S$. 
One has $(\rho,\vareps_{m-n+i})=(\rho,\delta_i)=1$ so
$s_{\delta_i}s_{\vareps_m+n-i}\rho=\rho+2\beta_i$ that is
$s_{\delta_i}s_{\vareps_m+n-i} Y=e^{\beta_i}Y$. 
Therefore
$$(1-s_{\delta_i})F(Y)=F(Y)+F(s_{\delta_i}s_{\vareps_m+n-i} Y)=
F\bigl((1+e^{\beta_i})Y\bigr)=F\bigl(\frac{e^{\rho+\beta_i}}{
\prod_{\beta\in S\setminus\{\beta_i\}}(1+e^{-\beta})}\bigr)=0,$$
because $s_{\vareps_m+n-i}\in W^{\#}$ stabilizes $\rho+\beta_i$ and
the elements of $S\setminus\{\beta_i\}$.
Thus $s_{\delta_i}F(Y)=F(Y)$ so $s_{\delta_i}s_{\delta_j}F(Y)=F(Y)$
for any $i,j$. Hence $X=F(Y)$ is $W_2$-skew-invariant.

%%%%%%%%%%%%%%%%  biblio.tex

\end{document}